\documentclass[sigplan,screen,authorversion]{acmart}

\setcopyright{acmcopyright}
\acmPrice{15.00}
\acmDOI{10.1145/3192366.3192388}
\acmYear{2018}
\copyrightyear{2018}
\acmISBN{978-1-4503-5698-5/18/06}
\acmConference[PLDI'18]{39th ACM SIGPLAN Conference on Programming Language Design and Implementation}{June 18--22, 2018}{Philadelphia, PA, USA}

\usepackage[utf8]{inputenc}
\usepackage[T1]{fontenc}
\usepackage{microtype}

\usepackage{epsfig,endnotes}
\usepackage{amsmath}
\usepackage{amssymb}
\usepackage{amsthm}
\usepackage{fancyvrb}
\usepackage[labelfont=bf]{caption}
\usepackage[caption=false]{subfig}
\usepackage{enumitem}
\usepackage{listings}
\usepackage{lstlinebgrd}
\usepackage{etoolbox}
\usepackage{color, colortbl}
\usepackage{pgfplots}
\usepackage{tikz}
\usetikzlibrary{calc}
\usetikzlibrary{patterns}
\usepackage{pifont}
\usepackage{multirow}
\usepackage{xspace}
\usepackage{url}
\usepackage{xr}
\usepackage{flushend}

\newcommand{\tool}[0]{Eff\-ect\-ive\-San\xspace}
\newcommand{\caver}[0]{\textsc{Ca\-Ver}\xspace}

\newcommand{\layoutF}[0]{\mathcal{L}}
\newcommand{\layout}[2]{\layoutF{}(#1, #2)}
\newcommand{\sizeofF}[0]{\mathsf{sizeof}}
\newcommand{\sizeof}[1]{\sizeofF(#1)}
\newcommand{\offsetofF}[0]{\mathsf{offsetof}}
\newcommand{\offsetof}[2]{\offsetofF(#1, #2)}
\newcommand{\set}[1]{\mathsf{#1}}
\newcommand{\pair}[2]{\langle #1, #2 \rangle}
\newcommand{\ignore}[1]{}
\newcommand{\mysubsubsection}[1]{\vspace{0.25em}\noindent\textbf{#1.}}
\newcommand{\base}[1]{\mathit{base}(#1)}
\newcommand{\size}[1]{\mathit{size}(#1)}
\newcommand{\cmark}[0]{\ding{51}}
\newcommand{\xmark}[0]{\ding{55}}

\definecolor{keyword}{rgb}{0.40, 0.40, 0}
\definecolor{comment}{rgb}{0, 0, 0.40}
\definecolor{type}{rgb}{0, 0.40, 0}
\definecolor{typecheck}{rgb}{0.40, 0, 0.40}
\definecolor{row}{rgb}{0.7,0.95,0.95}
\definecolor{row2}{rgb}{0.95,0.95,0.95}
\definecolor{row3}{rgb}{0.90,0.90,0.90}
\definecolor{mygreen}{rgb}{0,0.5,0}
\definecolor{mygreen2}{rgb}{0,0.25,0}

\newcommand{\type}[1]{\texttt{#1}}

\theoremstyle{definition}
\newtheorem{example}{Example}

\begin{document}

\date{}

\title[\tool: Dynamically Typed C/C++]
{\tool: Type and Memory Error Detection using Dynamically Typed C/C++}
\titlenote{
This research was partially supported by a grant from the
\grantsponsor{NRFsg}{National Research Foundation}{https://www.nrf.gov.sg/},
Prime Minister's Office, Singapore under its National
Cybersecurity R\&D Program (TSU\-NA\-Mi project, No.
\grantnum{NRFsg}{NRF2014NCR-NCR001-21})
and administered by the National Cybersecurity R\&D Directorate.
}

\author{Gregory J. Duck}
\affiliation{%
    \department{Department of Computer Science}
    \institution{National University of Singapore}
    \country{Singapore}
}
\email{gregory@comp.nus.edu.sg}
\author{Roland H. C. Yap}
\affiliation{%
    \department{Department of Computer Science}
    \institution{National University of Singapore}
    \country{Singapore}
}
\email{ryap@comp.nus.edu.sg}

\begin{abstract}
Low-level programming languages with weak/static type systems,
such as C and C++, are 
vulnerable to errors relating to the misuse of memory at runtime, such as
(sub-)object bounds overflows, (re)use-after-free,
and type confusion.
Such errors account for many security and other undefined behavior
bugs for programs written in these languages.
In this paper, we introduce the notion of \emph{dynamically typed C/C++},
which aims to detect such errors by dynamically checking the
``effective type''
of each object before use at runtime.
We also present an implementation of dynamically typed C/C++ in the form of
the \emph{Effective Type Sanitizer} (\tool).
\tool enforces type and memory safety using a combination of
low-fat pointers, type
meta data and type/bounds check instrumentation.
We evaluate 
\tool against the SPEC2006 benchmark suite and the Firefox web browser,
and detect several new type and memory errors.
We also show that \tool
achieves high compatibility and reasonable overheads for the
given error coverage.
Finally, we highlight that \tool is one of only a few tools that can 
detect \emph{sub-}object bounds errors, and uses a novel approach
(dynamic type checking) to do so.
\end{abstract}



\maketitle

\section{Introduction}\label{sec:intro}

Modern programming languages employ type systems 
to control object usage and detect bugs.
Type systems may be \emph{static} (compile time), \emph{dynamic} (run time),
\emph{strong} (strict), or \emph{weak} (loose).
The type system of C/C++ is static and weak, meaning that it is up
to the programmer to prevent type errors from occurring
at runtime, including:
bounds overflows 
(e.g., accessing the 101st element of an \type{int[100]}),
(re)use-after-free (type mutation) and type confusion
(e.g., bad cast) errors.
Detecting such errors is desirable for a number of reasons,
including: security, debugging,
conformance to the compiler's \emph{Type Based Alias Analysis}
(TBAA)~\cite{diwan98tbaa} assumptions,
C/C++ standards~\cite{c, cpp} compliance,
code quality (e.g., readability, portability, maintainability, etc.), and
revealing
type-related undefined behavior.
For example, type and memory errors are well known to be a major source of
security bugs,
e.g., accounting for over
75\% of remote code execution vulnerabilities in Microsoft
software alone~\cite{microsoft13sir}.
Type errors can also be problematic for reasons other than
security.
For example,
errors that violate the compiler's
TBAA assumptions may lead to
program mis-compilation---a known problem for
some SPEC2006 benchmarks~\cite{henning06spec}.

One solution
is to deploy a \emph{sanitizer} that instruments the
program with additional code aiming to detect errors at runtime.
Sanitizers are typically used for testing and debugging
during the development process---helping to uncover problems
before the software is deployed---and sometimes also for
hardening production code (with a performance penalty).
However, existing sanitizers tend to be specialized for specific classes of
errors rather than enforcing comprehensive dynamic type safety.
For example, Type\-San~\cite{haller16typesan},
Baggy\-Bounds~\cite{akritidis09baggy} and CETS~\cite{nagarakatte09softbound}
are specialized tools designed to detect type confusion,
bounds overflows and use-after-free errors respectively.
Neither tool offers any protection against other kinds of errors.
Furthermore, many sanitizers only offer incomplete protection against
the class of errors they target.
For example, a known limitation of
Add\-ress\-San\-it\-iz\-er~\cite{serebryany12asan},
Low\-Fat~\cite{duck16heap, duck17stack} and
Baggy\-Bounds~\cite{akritidis09baggy}
is that they do not protect against sub-object bounds overflows, e.g.:
\begin{Verbatim}
struct account {int number[8]; float balance;}
\end{Verbatim}
Modification of (\texttt{balance}) from an overflow in the account
(\texttt{number}) will not be detected.
Another example is \caver~\cite{lee15caver}, TypeSan~\cite{haller16typesan},
and HexType~\cite{jeon17hextype} which are specialized to detect bad
casts between C++ \texttt{class} types only.

In this paper, we propose dynamic type checking for C/C++ as a unified
method for detecting
a wide range of memory misuse errors,
including type confusion, (sub-)object bounds overflows and
(re)use-after-free.
We also propose an implementation of dynamic type
checking in the form of the \emph{Effective Type Sanitizer} (a.k.a. \tool).
\tool dynamically verifies the \emph{effective} type
(see \cite{c} \S 6.5.0 \P 6) of each object before use,
allowing for the direct detection of the following classes of errors:
\begin{itemize}[leftmargin=*]
\item \underline{\textbf{Type-errors}}:
By dynamically checking types, \tool directly detects
type errors that are a common source of
security vulnerabilities (e.g., type confusion~\cite{lee15caver})
and other undefined behavior.
\tool's type checking is comprehensive, covering all standard C/C++ types
(\type{int}, \type{float}, pointers, \texttt{struct}s, \texttt{class}es,
\texttt{union}s, etc.).
Furthermore, coverage is not limited to explicit cast operations.

\item \underline{\textbf{(Sub-)Object-bounds-overflows}}:
C/C++ types intrinsically encode bounds information,
e.g., (\type{int[100]}),
meaning that type and bounds checking go hand-in-hand.
\tool uses dynamic types to detect bounds errors, as well as
sub-object bounds overflows within the same object.
\end{itemize}
Furthermore,
\tool can detect some classes of
(re)use-after-free errors: 
\begin{itemize}[leftmargin=*]
\item \underline{\textbf{(Re)Use-after-free}} and
    \underline{\textbf{Double-free}}:
By binding unallocated objects to a special type, \tool can also
detect some use-after-free and double-free errors.
Likewise, reuse-after-free
(when the object is reallocated before the erroneous access)
is protected if the reallocated object is bound to a different type.
\end{itemize}
In essence, \tool is a ``generalist'' sanitizer 
that finds multiple classes of errors using a 
single underlying methodology, namely,
dynamic type checking.
Such errors account for the majority of 
attacks~\cite{microsoft13sir} as well as other undefined behavior.
Furthermore, unlike existing C/C++ type error sanitizers~\cite{
haller16typesan, lee15caver, jeon17hextype, kell16type},
\tool checks pointer \emph{use} (i.e., dereference) rather than explicit casts.

Our \tool implementation works by extending
low-fat pointers~\cite{duck16heap, duck17stack} to
dynamically bind type meta data to allocated objects.
Low-fat pointers have several advantages, including: speed, low 
memory overheads and compatibility with uninstrumented code.
The key insight is to store meta data at the base of allocated objects,
analogous to a hidden \verb+malloc+ header,
which can be retrieved using standard low-fat pointer operations.
This differs from most existing sanitizers that store meta data
in a shadow space or some other adjunct memory.
\tool's type meta data is detailed, storing the type and bounds of every
possible sub-object,
allowing for \emph{interior pointers} (pointers to sub-objects inside
allocated objects) to be checked at runtime.
We experimentally evaluate \tool against the SPEC2006 benchmarks and
the Firefox web browser~\cite{firefox}.
\tool finds multiple type, (sub-)object bounds,
and reuse-after-free errors in SPEC2006,
with some errors previously unreported.

\tool offers more comprehensive error detection compared to more specialized
tools.
However, more comprehensive error detection necessitates more instrumented
checks, so the trade-off is higher performance overheads.
\tool is intended for deployment in the
software development and testing life-cycle where error coverage is
the priority.
While \tool's design philosophy is to ``check everything'' by default,
it is also possible to trade coverage for performance.
To demonstrate this, we also evaluate
two reduced-instrumentation variants of \tool, namely:
\begin{itemize}[leftmargin=*]
\item \tool-type: for type-cast-checking-only; and
\item \tool-bounds: for bounds-checking-only.
\end{itemize}
Both variants have similar coverage compared to
existing state-of-the-art specialized sanitizers.
In summary, the main contributions of this paper are:
\begin{itemize}[leftmargin=*]
\item[-] \emph{Dynamic Type Checking}: We introduce dynamically
    typ\-ed C/C++ as a general methodology against a wide range of errors
    relating to the misuse of memory.
\item[-] \emph{\tool}: 
    We present a practical implementation of dynamically typed C/C++ in
    the form of the \emph{Effective Type Sanitizer} (\tool).
    \tool offers comprehensive type error detection
    (for both C and C++),
    comprehensive (sub-)object bounds overflow error detection,
    as well as partial detection for some
    (re-)use-after-free errors,
    all using the same underlying methodology.
\item[-] \emph{Sub-object Bounds Checking}:
    Dynamic type checking offers a novel approach to sub-object bounds
    checking.
    Most existing bounds-checking tools either check object bounds only
    (e.g., AddressSanitizer~\cite{serebryany12asan}), or require
    explicit tracking of sub-object bounds information, e.g., by changing the
    \emph{Application Binary Interface} (ABI)
    (e.g., SoftBound~\cite{nagarakatte09softbound}), however this can be
    a source of incompatibility.
    In contrast, \tool uses dynamic type information to derive sub-object
    bounds ``on demand'',
    does not change the ABI, and is thread-safe.
\item[-] \emph{Evaluation}: We experimentally evaluate \tool
    ag\-ainst the SPEC2006 benchmark suite~\cite{henning06spec} and the
    Firefox web browser~\cite{firefox}.
    SPEC2006 is a heavily analyzed code-base, yet \tool is able
    to detect several new errors.
\end{itemize}

\section{Background}\label{sec:background}

Dynamically typed languages, such as JavaScript, Python, Lua, etc.,
check the types of objects at runtime.
In contrast, statically typed languages, such as C, check
types at compile time.
Similarly, C++ is a statically typed language with the limited exception of
\emph{Run-Time Type Information} (RTTI) and 
the (\verb+dynamic_cast+) operator for downcasting
(casting from a base to a derived class).
The C/C++ type system is 
intentionally weak, i.e., allowing for arbitrary pointer casting
and pointer arithmetic,
meaning that type and memory errors will not be prevented at
compile time.
By using dynamic typing, we can detect such errors at runtime at the
cost of additional overheads.
Note that
dynamic typing concerns pointer or reference access
only, e.g., (\texttt{f = *(float *)p}) is a type error if
\texttt{p} does not point to a (\type{float}) object.
Casts that create copies of objects, such as
(\texttt{f = (float)i}), are valid conversions and not type errors.

Aside from RTTI,
there is limited existing work on dynamic type checking for C/C++.
A simple dynamic checking system for C that tags each data word with 
a basic type, e.g., \emph{integral}, \emph{real}, \emph{pointer}, etc.,
was proposed in \cite{loginov01typecheck}.
Unlike our approach, there is no distinction between different types of
pointers (i.e., all pointers are treated as (\type{void *})).
Overheads are also very high at 35$\times$-133$\times$ for
SPEC95~\cite{loginov01typecheck}.
Type confusion sanitizers also
provide a limited form of dynamic typing discussed below.
Bounds-checking is also a weak form of dynamic typing
(where only the type's size is checked).
CCured~\cite{necula05ccured} extends the C type system with memory-safety
guarantees.
However, CCured has limited compatibility, no C++ support, and does not
track types over arbitrary casts.

\subsection{Sanitizers}\label{sec:tools}

Type and memory errors have long been recognized as a
major source of bugs in programs
written in low-level languages such as C/C++.
As such, many different bug detection tools (sanitizers) have been proposed
which we survey and compare. 

\mysubsubsection{Type Confusion}
C++ provides a limited form of dynamic typing in the
form of RTTI and (\verb+dynamic_cast+) for downcasting.
However, programmers will sometimes opt for the
faster yet unsafe (\verb+static_cast+) version of the same
op\-er\-at\-ion---a known source of security vulnerabilities.
\caver~\cite{lee15caver} and TypeSan~\cite{haller16typesan} 
are specialized for detecting such type confusion errors caused by
unsafe downcasts.
Another approach is the \emph{Undefined Behavior Sanitizer}
(UBSan)~\cite{ubsan} that transforms
\verb+static_cast+s into \verb+dynamic_cast+s to enable standard
RTTI protections.
HexType~\cite{jeon17hextype} is a more general tool that extends
protection to other kinds of C++ casts such as
(\verb+reinterpret_cast+), (\verb+const_cast+), etc.
Finally, \textsf{libcrunch}~\cite{kell16type} can 
detect bad pointer casts for C programs.

Existing type confusion sanitizers have several limitations.
Firstly, existing sanitizers only verify \emph{incomplete}
types that lack bounds information (e.g., \type{$T$[]} is \emph{incomplete}
whereas \type{$T$[100]} is \emph{complete}).
For example, if (\verb+p+) points to an object of
type (\type{$T$[100]}), then (\verb_p+101_) may point to an object of
any type or unused memory.
Existing sanitizers will not detect such bounds errors since they assume both
(\verb+p+) and (\verb_p+101_) have the same type (\type{$T$[]}).
\tool detects (sub-)object bounds errors based on
complete type information.
The second limitation is that existing sanitizers
instrument explicit cast operations only.
\emph{Implicit casts}
(e.g., via memory, unions, function arguments, etc.) are unprotected.
For example, the following is an
implicit cast from (\verb_ptrA_) to (\verb_ptrB_):
\begin{Verbatim}
 memcpy(buf, &ptrA, 8); memcpy(&ptrB, buf, 8);
\end{Verbatim}
\tool instruments \emph{pointer use} (i.e., dereference) meaning that
type errors arising from (\verb+ptrB+)'s usage will be detected
(regardless of how the cast occurred).
The final limitation is that existing sanitizers focus only on a
subset of explicit C/C++ casts, e.g.,
C++ \texttt{class} casts for \caver/Type\-San/HexType.
In contrast, \tool can detect type errors for any C/C++ type
(\type{int}, \type{float}, \verb+struct+s, pointers, etc.).
\tool generally performs more type checks than existing tools,
mainly because of increased type coverage
and pointer use instrumentation.

\mysubsubsection{(Sub-)Object Bounds Overflows}
Object bounds overflows are well known to be a major source of
security vulnerabilities.
As such, many existing solutions have been proposed,
including~\cite{akritidis09baggy, austin94safec, duck16heap, duck17stack,
    eigler03mudflap, mpx, jim02cyclone, kwon13lowfat, nagarakatte09softbound,
    necula05ccured, serebryany12asan, younan10paricheck}
amongst others.
Many solutions, such as BaggyBounds~\cite{akritidis09baggy},
LowFat~\cite{duck16heap, duck17stack},
Intel MPX~\cite{mpx} and SoftBound~\cite{nagarakatte09softbound},
work by binding \emph{bounds meta data} (object size and base) to
each pointer.
The binding is typically implemented using some form of shadow memory
(e.g., SoftBound, MPX) or encoding the meta data within the pointer
itself (e.g., LowFat with low-fat pointers).
Solutions that use shadow memory may also have compatibility issues
interfacing with uninstrumented code that allocates its own memory
(the corresponding entries in the shadow memory will not be initialized).
This can be partially mitigated by intercepting standard memory allocation
functions, or by hardware-based solutions (e.g., with MPX).
Low-fat pointers avoid the problem by encoding bounds meta data within the
pointer itself.
Another approach to memory safety is AddressSanitizer~\cite{serebryany12asan} 
which uses \emph{poisoned red-zones} and shadow memory to track the state of
each word of memory, e.g.
\emph{unallocated}, \emph{allocated} or \emph{red-zone}.
Out-of-bounds
memory access that maps to a red-zone will be detected, however, memory errors
that ``skip'' red-zones may be missed.

Most existing bounds overflow sanitizers protect
\emph{allocation} or \emph{object bounds only}.
This means the overflows contained within an allocated object will not
be detected, e.g., the overflow into (\texttt{balance}) from
Section~\ref{sec:intro}.
A few bounds checking systems, 
e.g., SoftBound~\cite{nagarakatte09softbound}
and Intel MPX~\cite{mpx}, can also detect sub-object bounds overflows
by using static type information for \emph{bounds narrowing}, i.e.,
an operation that further constrains
bounds meta information to a specific sub-object
for more accurate protection.
This also requires sub-object bounds to be associated with pointers
when they are passed between contexts, e.g., a pointer parameter in a
function call.
For example, MPX solves this problem by passing bounds information through
special registers (\texttt{bnd0}-\texttt{bnd3}), or failing that,
by resorting to the \emph{bounds directory} stored in shadow memory.
Similarly, SoftBound also explicitly tracks bounds information, e.g.,
by inserting additional function parameters~\cite{nagarakatte09softbound}.
Both MPX and SoftBound use shadow memory schemes that have been shown
to be unsuitable for multi-threaded environments~\cite{oleksenko17mpx}.
In contrast, \tool detects (sub-)object bounds errors using dynamic type
information.
For example, a pointer (\verb+p+) of static type (\type{int *}) can be
matched against an object of dynamic type (\type{account}), since
(\verb+p+) points to the sub-object (\texttt{number}) of compatible type.
\tool will enforce the sub-bounds for (\texttt{number}), thereby
preventing overflows into (\texttt{balance}) or outside the (\texttt{account}).
Unlike other sub-object bounds checkers, \tool does not change the
\emph{Application Binary Interface} nor rely on thread-unsafe shared state.

\mysubsubsection{(Re)Use-After-Free}
Use-after-free (UAF) sanitizers include tools such as
AddressSanitizer~\cite{serebryany12asan} and Compiler Enforced Temporal Safety
(CETS)~\cite{nagarakatte10cets}.
AddressSanitizer stores the allocation state in shadow memory, allowing
for the detection of use-after-free errors.
AddressSanitizer also mitigates reuse-after-free by putting
freed objects into a ``quarantine'' that delays reallocation
(a technique also applicable to \tool).
CETS uses a more sophisticated identifier-based approach, that binds a unique
tag to each allocated object, allowing for general
(re)use-after-free detection. 

\tool's use-after-free protection is related to the
AddressSanitizer approach---but with
type meta data replacing AddressSanitizer's shadow memory scheme.
\tool can also detect reuse-after-free provided the object is
reallocated with a different type.
Although \tool's protection is not as
comprehensive as specialized tools such as CETS,
it is nevertheless worthwhile to target such errors anyway,
since this incurs no additional costs.
Tools based on instrumentation
(including \tool, CETS, AddressSanitizer)
may also miss some use-after-free errors because of the
inherent race between the check and a call to (\texttt{free}),
e.g., by another thread.
(Re)use-after-free can also be mitigated using other
means, such as garbage collection~\cite{boehm88gc}.

\begin{figure}
\centering
{\small
\begin{tabular}{|l|ccc|}
\hline
Sanitizer & {\scriptsize Types} & {\scriptsize Bounds} &
    {\scriptsize UAF} \\
\hline
\hline
\caver~\cite{lee15caver} 
                 & Partial$^*$  & \xmark & \xmark \\
TypeSan~\cite{haller16typesan}
                 & Partial$^*$  & \xmark & \xmark \\
UBSan~\cite{ubsan}
                 & Partial$^*$ & \xmark & \xmark \\
HexType~\cite{jeon17hextype}
                 & Partial$^*$ & \xmark & \xmark \\
\textsf{libcrunch}~\cite{kell16type}
                 & Partial$^\text{\textasciicircum}$ & \xmark & \xmark \\
\hline
BaggyBounds~\cite{akritidis09baggy} 
                 & \xmark & Partial$^\dagger$ & \xmark  \\
LowFat~\cite{duck16heap, duck17stack}
                 & \xmark & Partial$^\dagger$ & \xmark  \\
Intel MPX~\cite{mpx}
                 & \xmark & \cmark            & \xmark  \\
SoftBound~\cite{nagarakatte09softbound}
                 & \xmark & \cmark            & \xmark \\
\hline
CETS~\cite{nagarakatte10cets}
                & \xmark & \xmark            & \cmark \\
\hline
AddressSanitizer~\cite{serebryany12asan} 
                 & \xmark & Partial$^\dagger$ & Partial$^\ddagger$ \\
SoftBound+CETS~\cite{nagarakatte09softbound, nagarakatte10cets}
                 & \xmark      & \cmark & \cmark \\
\hline
\hline
\textbf{\tool}   & \cmark      & \cmark & Partial$^\S$ \\
\hline
\end{tabular}
}
\caption{Summary of different sanitizers and capabilities against
type and memory errors.
Here 
(\cmark) means comprehensive protection,
(\xmark) means no or incidental protection,
and (Partial) means partial protection with caveats.
The caveats are: ($*$) only protects a subset of explicit C++ casts,
($\text{\textasciicircum}$) only protects explicit C casts,
($\dagger$) only protects allocation bounds,
($\ddagger$) only protects use-after-free (not reuse-after-free),
and ($\S$) only protects reuse-after-free for different types.
\label{fig:tools}}
\vspace{-4mm}
\end{figure}

\subsection{Our Approach}
Figure~\ref{fig:tools} summarizes existing sanitizers and their
capabilities.
Most sanitizers are specialized to one particular class of error
and/or offer partial protection against the classes of errors they
do support.
This means that, if more comprehensive error detection is desired, 
multiple different tools must be deployed at once.
However, this is problematic, since
most sanitizers are compiler specific
(e.g., \texttt{clang} versus \texttt{gcc}) and
use competing in\-stru\-ment\-at\-ion/sha\-dow-mem\-ory
schemes that are not generally designed to be interoperable.
Even if it were possible to seamlessly combine sanitizers,
\tool still offers a more comprehensive level of error detection,
such as type errors caused by implicit casts.

\tool's underlying approach is to convert C/C++ into a dynamically
typed programming language.
The basic idea is to bind a \emph{dynamic type} to each allocated object,
which can be retrieved at runtime and compared against
the static type declared by the programmer.
The dynamic type information is complete and
supports standard C/C++ types, thus allowing for the detection of type errors
beyond
\caver, TypeSan, HexType and \textsf{libcrunch}.
Furthermore, C/C++ types encode (sub-)object size information, and
thus dynamic types can be used to enforce (sub-)object bounds.
\tool's bounds enforcement is precise and 
offers more comprehensive error detection than 
BaggyBounds, LowFat and AddressSanitizer.
Finally, by binding deallocated objects to a special type, dynamic
typing can also detect some (re)use-after-free errors.
Although use-after-free detection is partial,
it incurs no additional costs
while still detecting many common cases.

\section{Dynamic Types for C/C++}\label{sec:types}

In this section, we present a dynamic type system for C/C++.
This is essentially equivalent to the standard (static) type
system, but also includes extensions for handling unallocated memory,
and methods for calculating sub-object types and bounds at runtime.

The dynamic type of an object is a
qualifier-free\footnote{Qualifiers do not affect memory layout
or access (\cite{c} \S 6.5.0 \P 7).}
version of the \emph{effective type} (\cite{c} \S 6.5.0 \P 6) 
or \emph{object type} (\cite{cpp} \S 3.9.0 \P 8) as
defined by the C/C++ standards.
The dynamic type can be any C/C++ type, including
fundamental types (e.g., \type{int}, \type{float}, etc.),
pointers, function pointers, arrays, structures, classes and unions.
Dynamic types are always \emph{complete}, 
i.e., the type's size is known.
We assume w.l.o.g. that type aliases (e.g., \type{typedef})
are fully expanded and
C++ templates and namespaces are fully instantiated.
Structures, classes and unions are considered equivalent based on tag
(i.e., the name),
or in the case of anonymous types, based on layout.
We denote the set of all types as ($\set{Type}$).
For brevity, we use the C++ convention of referring to types by their tag,
e.g., (\type{S}) is short for (\type{struct S}).

During a new allocation (e.g., stack allocation or heap allocation via
\texttt{malloc}, \texttt{new}, \texttt{new[]})
the dynamic type will be bound to the object.
For stack allocations and C++'s \texttt{new}/\texttt{new[]} operators,
the dynamic type is the same as the declared type of the object
as defined by the program.
For \texttt{malloc}
the dynamic type is deem\-ed equivalent to
the first lvalue usage type.
The latter is determined by a simple program analysis.

\begin{example}[Dynamic Types]\label{ex:alloc_types}
Consider the type definitions and allocations:
\begin{Verbatim}[samepage]
struct S {int a[3]; char *s;};
struct T {float f; struct S t;};
S x[8]; q = new T; r = (T *)malloc(sizeof(T));
s = (T *)malloc(100*sizeof(T));
\end{Verbatim}
Pointer \texttt{x} will be bound to type (\type{S[8]}),
\texttt{q} and \texttt{r} bound to type
(\type{T[1]}), and \texttt{s} bound to type
(\type{T[100]}).
Notice that all dynamic types are \emph{complete}, where the type's size is
determined by the allocation size.
$\qed$\end{example}

\begin{figure}
\small
\begin{align*}
\renewcommand*{\arraystretch}{1.1}
\begin{array}{l|rl|}
\cline{2-3}
&
\layoutF & : \set{Type} \times \mathbb{Z} \mapsto
    P(\set{Type} \times \mathbb{Z}) \\
\cline{2-3}
\text{\small (a)} &
\layout{T}{0} & \ni \langle T, 0 \rangle \\
\text{\small (b)} &
\layout{T}{\sizeof{T}} & \ni \langle T, \sizeof{T} \rangle \\
\text{\small (c)} &
\layout{\texttt{$T$[$N$]}}{k} & \supseteq 
    \layout{$T$}{k \bmod \sizeof{T}} \\
\text{\small (d)} &
\layout{\texttt{$T$[$N$]}}{k} & \ni
    \langle \texttt{$T$[$N$]}, k \rangle 
    ~\text{if}~k \bmod \sizeof{T} = 0 \\
\text{\small (e)} &
\layout{\texttt{struct S}}{k} & \supseteq
    \layout{T_\mathit{memb}}{k -
        \offsetof{\texttt{S}}{\mathit{memb})}} \\
\text{\small (f)} &
\layout{\texttt{class C}}{k} & \supseteq
    \layout{T_\mathit{memb}}{k -
        \offsetof{\texttt{C}}{\mathit{memb})}} \\
\text{\small (g)} &
\layout{\texttt{union U}}{k} & \supseteq
    \layout{T_\mathit{memb}}{k} \\
\text{\small (h)} &
\layout{\texttt{FREE}}{k} & =
    \{\langle \texttt{FREE}, 0 \rangle \} \\
\cline{2-3}
\end{array}
\end{align*}
\vspace{-1em}
\caption{\label{fig:layout}The layout function ($\layoutF$).
    Rules (c)-(h) implicitly assume that $k$ is within the bounds of the
    object, that is, $0 {\leq} k {<} \sizeof{T}$ for rule matching
    $\layout{T}{k}$.
    Rules (e)-(g) apply to all members ($\mathit{memb}$) of the
    corresponding structure/class/union.
    Here ($\sizeofF$) and ($\offsetofF$) are the standard ANSI C operators.
}
\end{figure}

\mysubsubsection{Deriving Sub-object Types}
The dynamic type represents the type of the top-level allocated object.
In C/C++, it is common for pointers to point to sub-objects contained within
larger objects---so called \emph{interior} pointers.
Interior pointers can point to array elements, or to
a member contained within a structure, class or union.
Another example is C++ classes with inheritance, where
base class(es) are typically implemented as sub-objects of the
derived class.

\tool 
explicitly tracks the dynamic type of top-level allocated objects.
Sub-object types are derived dynamically 
from the containing allocated object's type
(or containing type for short)
and an \emph{offset}, i.e.,
the pointer difference (in bytes) between the interior pointer and the base
pointer of the containing object.
For the ease of presentation, we shall assume all pointer
arithmetic uses byte offsets regardless of the underlying pointer type.
To derive sub-object types,
we assume a runtime system that can map interior
pointers to containing types and offsets
(see Section~\ref{sec:runtime}).
Next the containing type and offset is mapped to the set of possible
sub-object types using
a \emph{memory layout function}, denoted ($\layoutF$),
that is formalized
as the relation defined inductively over rules (a)-(h) from
Figure~\ref{fig:layout}.
Essentially,
given a pointer $p$ to the base of an allocated object of dynamic type $T$
and an offset $k$ (in bytes), the function
$\layout{T}{k}$ returns the set of type/integer pairs $\pair{U}{\delta}$
that represent
{\it all valid sub-objects} pointed to by pointer $(p{+}k)$.
Here, the type (\type{$U$}) represents the sub-object's type, and
integer $\delta$ represents the distance from the pointer
$(p{+}k)$ to the sub-object's base.
The integer $\delta$ is used later for sub-object bounds calculation.
For example, the layout for \type{int} assuming $\sizeof{\type{int}}{=}4$ is
{
\setlength\abovedisplayskip{4pt}
\setlength\belowdisplayskip{4pt}
\begin{align*}
\begin{array}{l}
\layout{\type{int}}{0} = \{\pair{\type{int}}{0}\} \text{\hspace{1cm}}
\layout{\type{int}}{4} = \{\pair{\type{int}}{4}\} \\
\layout{\type{int}}{k} = \emptyset \text{\hspace{0.5cm}(otherwise)}
\end{array}
\end{align*}
}%
Thus, if $p$ points to \type{int}, then both $(p{+}0)$
and $(p{+}4)$
also point to \type{int} by rules Figure~\ref{fig:layout}(a)-(b)
respectively.
Rule (b) accounts for the one-past-the-last-element required
by the C standard (\cite{c} \S 6.5.6 \P 7,8).
The layouts for other fundamental types, pointers, functions and enumerations
are defined similarly.
For compound types (arrays, structures and unions) we 
build more complicated layouts.
Rules~(e)-(g) state that the layout of a struct/class/union
member ($\mathit{memb}$) of type ($T_\mathit{memb}$) includes
the layout of ($T_\mathit{memb}$) offset within the containing type
(the offset is zero for unions).
Similarly rule (c) for arrays.
For classes with inheritance, we consider any base class to be an implicit
embedded member.
Finally, special rule~(d) states that interior pointers
to array elements can also be considered pointers to the containing array
itself.
This is necessary because a common idiom is to scan arrays using pointers
rather than element indices.
\begin{example}[Structure Layout]\label{ex:layout}
Consider a pointer $p$ to type (\type{T}) defined in
Example~\ref{ex:alloc_types}.
Then all (sub-)objects for $p$ are described by the following
table:
\begin{center}
\setlength{\tabcolsep}{2pt}
\begin{tabular}{l|c|c}
Sub-obj. & Offset & Type \\
\hline
\hline
$p$                  & $p{+}0$  & \type{T} \\
$p\texttt{->f}$      & $p{+}0$  & \type{float} \\
$p\texttt{->t}$      & $p{+}4$  & \type{S} \\
$p\texttt{->t.a}$    & $p{+}4$  & \type{int[3]} \\
\end{tabular}
\hspace{0.5cm}
\begin{tabular}{l|c|c}
Sub-obj. & Offset & Type \\
\hline
\hline
$p\texttt{->t.a[0]}$ & $p{+}4$  & \type{int} \\
$p\texttt{->t.a[1]}$ & $p{+}8$  & \type{int} \\
$p\texttt{->t.a[2]}$ & $p{+}12$ & \type{int} \\
$p\texttt{->t.s}$    & $p{+}16$ & \type{char *} \\
\end{tabular}
\end{center}
Consider $(p{+}4)$, which points to the base of sub-objects
($p\texttt{->t}$), ($p\texttt{->t.a}$) and ($p\texttt{->t.a[0]}$)
respectively, as well as
pointing to the end of sub-object ($p\texttt{->f}$).
Using the rules from Figure~\ref{fig:layout}, we derive:
{
\setlength\abovedisplayskip{4pt}
\setlength\belowdisplayskip{4pt}
\begin{align*}
\layout{\type{T}}{4} = \{
    \pair{\type{S}}{0}, \pair{\type{int[3]}}{0}, \pair{\type{int}}{0},
    \pair{\type{float}}{4}
\}
\end{align*}
}%
Pointers to array elements can also be treated as pointers to
the array itself (rule Figure~\ref{fig:layout}(c)).
Thus for $(p{+}12)$:
{
\setlength\abovedisplayskip{4pt}
\setlength\belowdisplayskip{4pt}
\begin{align*}
\layout{\type{T}}{12} = \{
    \pair{\type{int[3]}}{8}, \pair{\type{int}}{0}, \pair{\type{int}}{4}
\} 
\end{align*}
}%
corresponding to the array sub-object ($p\texttt{->t.a}$) via
rule~(c), the array element ($p\texttt{->t.a[2]}$) and
the end of the previous array element ($p\texttt{->t.a[1]}$),
respectively.
$\qed$\end{example}\noindent
The layout for compound objects is a \emph{flattened} representation,
meaning that ($\layoutF$) returns types for even nested objects,
e.g., ($p\texttt{->t.a[2]}$) is three levels deep.

Finally, we remark that \texttt{union}s (\texttt{U}) are treated no
differently to \texttt{struct}s or \texttt{class}es, except that
the member offset is defined to be zero,
i.e., ($\offsetof{\texttt{U}}{\mathit{memb}} = 0$).
This means that members always overlap.
However, even \texttt{struct}s may have overlapping sub-objects,
as demonstrated by Example~\ref{ex:layout}.

\mysubsubsection{A Special Type for Deallocated Memory}
Deallocated objects are bound to a special type
(\type{FREE}) that is defined to be distinct from all other C/C++ types.
This reduces use-after-free and double-free errors to type errors
without any other special treatment.
The (\type{FREE}) type has a special layout defined by rule
Figure~\ref{fig:layout}(h).
Essentially, if $p$ points to deallocated memory, then so does
$(p{+}k)$ for all $k$.
Reuse-after-free is already handled for the case where
the reallocated object has a different type to that of the dangling pointer.

\mysubsubsection{Calculating Sub-object Bounds}
C/C++ types also encode bounds information.
For example, the type (\type{int[100]}) is an array object of length 100,
and accessing an element outside the range $0..99$ is an object bounds error.
Hence, full dynamic type checking necessitates the enforcement of object bounds
at runtime.
To support this, we calculate (sub-)object bounds 
using dynamic type information.
The basic idea is as follows:
let $p$ point to an object of type $T$ and $q{=}(p{+}k)$, then each
pair $\pair{U}{\delta} \in \layout{T}{k}$ corresponds to a sub-object
of type $U$ pointed to by $q$.
The integer $\delta$ represents the distance from $q$ to the start of
the sub-object, and is not necessarily zero
(e.g., interior pointers to arrays).
The sub-object bounds, represented as an address range,
can be calculated using the following helper function:
{
    \setlength\abovedisplayskip{4pt}
    \setlength\belowdisplayskip{4pt}
\begin{align*}
    \mathtt{type\_bounds}(q, \pair{U}{\delta}) =
    q{-}\delta~..~q{-}\delta{+}\sizeof{U}
\end{align*}
}%
For example, let us consider the pointer $q{=}(p{+}12)$ into the sub-object
($p\texttt{->t.a}$) corresponding to the pair
$\pair{\type{int[3]}}{8} \in \layout{\type{T}}{12}$ from
Example~\ref{ex:layout}.
The sub-object bounds for ($p\texttt{->t.a}$) is $(p{+}4)..(p{+}16)$,
i.e., spanning offsets $4..16$ bytes.

\section{Dynamic Type Check Instrumentation}\label{sec:instrumentation}

The aim of \emph{dynamic type checking} is to verify that pointer \emph{use}
(a.k.a., pointer dereference) is consistent with the
dynamic type of the underlying object.
The basic idea is as follows:
suppose pointer $p$ with static type (\type{$T$ *}) is dereferenced,
then dynamic type checking
verifies the following properties:
\begin{itemize}[leftmargin=*]
\item[-] \emph{Type Correctness}: pointer $p$ must point to the
    $i^\mathit{th}$ element of an
    object with dynamic type (\type{$T$[$N$]}) for some $i,N$; and
\item[-] \emph{Bounds Correctness}: index $i$ must be within the bounds of
    the object, i.e., $i \in 0{..}N{-}1$.
\end{itemize}
These properties ensure that the dereference is consistent with
the complete dynamic type (\type{$T$[$N$]})---including both the incomplete
type (\type{$T$[]}) and bound ($N$)---effectively transforming C/C++ into a
dynamically typed programming language.

\tool implements dynamic type checking in the form of
\emph{dynamic type check instrumentation} which
ensures that all pointer use is guarded by
an explicit type/bounds check.
For performance reasons, \tool's instrumentation also aims to minimize the
number of type checks.
One key observation is that the dynamic type is invariant
w.r.t. pointer arithmetic, e.g., for $q{=}p{+}k$, then pointers $p$ and $q$
reference the same underlying object, and hence the same type.
Thus, only $p$ need be type checked provided the derived pointer $q$ remains
within the bounds of the object.
Similarly, we can avoid type checking field access
$q{=}\texttt{\&}p\texttt{->}m$.
\tool's dynamic type check instrumentation schema is shown in
Figure~\ref{fig:schema}, and is summarized as follows:
\begin{itemize}[leftmargin=*]
\item[-] Figure~\ref{fig:schema}(a)-(d): \emph{Type checking input pointers}.
    All \emph{input pointers} (i.e., function parameters \ref{fig:schema}(a),
	call returns \ref{fig:schema}(b),
    pointers read from memory \ref{fig:schema}(c) and
	pointers created by casts \ref{fig:schema}(d)\footnote{
        For our purposes, we consider pointers created by casts to be inputs.
	})
    are type check\-ed against the incomplete\footnote{
As a simplification, we assume that all static types are incomplete.
A complete static type check can be decomposed into an incomplete type check
followed by a bounds narrowing operation.}
    static type declared by the programmer.
    The check also calculates the (sub-)object bounds
    based on the dynamic type, representing the address range
    for which the static type is correct.
\item[-] Figure~\ref{fig:schema}(e)-(f): \emph{Propagating/narrowing
    bounds to derived pointers}.
    Rule~\ref{fig:schema}(e) covers pointer arithmetic and
    \ref{fig:schema}(f) field access; and
\item[-] Figure~\ref{fig:schema}(g): \emph{Bounds checking}
	all pointer use/escapes.
\end{itemize}
Rule~(d) also extends to other kinds of casts such
as integer-to-pointer, C++'s \verb+static_cast+s, etc.
The type check is implemented as a call to a
special (\verb+type_check+) function supplied by the
\tool runtime system (to be defined later in Section~\ref{sec:runtime}).
The (\verb+type_check+) function will
log an error message if the pointer
does not point to an object, or sub-object of a larger object, of
the complete type (\type{type[$N$]}) for some $N$.
For example:
\begin{Verbatim}[samepage]
    int *p = new int[100];
    BOUNDS b1 = type_check(p, int[]);
    BOUNDS b2 = type_check(p, float[]);
\end{Verbatim}
The first type check passes but the second fails since (\type{int}) and
(\type{float}) are distinct types.
Assuming there is no error, the (\verb+type_check+) function will also
return the bounds of the matching (sub-)object.
Bounds are represented by a pair of pointers, e.g.,
$\texttt{b1}{=}\{p .. p+100*\sizeof{\type{int}}\}$.

\begin{figure}[t]
{\small
\newcommand{\highlightex}[1]{%
    \ifboolexpr{
        test {\ifnumcomp{#1}{=}{2}}
    }{\color{black!15}}{}
}
\newcommand{\highlightexx}[1]{%
    \ifboolexpr{
        test {\ifnumcomp{#1}{=}{1}}
    }{\color{black!15}}{}
}
\lstset{language=C,
                basicstyle=\ttfamily\fontseries{m}\selectfont,
                deletekeywords={int},
                keywordstyle=\color{keyword}\ttfamily\fontseries{b}\selectfont,
                keywordstyle=[2]\color{type}\ttfamily\fontseries{b}\selectfont,
                keywordstyle=[3]\color{typecheck}\ttfamily\fontseries{b}\selectfont,
                stringstyle=\color{red}\ttfamily,
                commentstyle=\color{comment}\ttfamily\fontseries{b}\selectfont,
                morekeywords=[2]{void, TYPE, META, SUBOBJECT, BOUNDS, ssize_t,
                    size_t},
                morekeywords=[3]{type_check, type_match, type_error,
                    type_bounds, type_lookup, bounds_narrow, malloc,
                    lowfat_malloc, typed_malloc, bounds_check},
                linebackgroundcolor=\highlightex{\value{lstnumber}},
                escapechar=@,
                mathescape
}
\setlength{\tabcolsep}{2pt}
\begin{tabular}{rl}
{(a)} &
\begin{minipage}{0.93\columnwidth}
\begin{lstlisting}
f(type *p) {
BOUNDS b = type_check(p, @\textcolor{type}{type[]}@);
$\cdots$ }
\end{lstlisting}
\end{minipage} \\
\hline
{(b)} &
\begin{minipage}{0.93\columnwidth}
\begin{lstlisting}
type *p = f($\cdots$);
BOUNDS b = type_check(p, @\textcolor{type}{type[]}@);
\end{lstlisting}
\end{minipage} \\
\hline
{(c)} &
\begin{minipage}{0.93\columnwidth}
\begin{lstlisting}
type *p = *q; 
BOUNDS b = type_check(p, @\textcolor{type}{type[]}@);
\end{lstlisting}
\end{minipage} \\
\hline
{(d)} &
\begin{minipage}{0.93\columnwidth}
\begin{lstlisting}
type *p = (type *)q; 
BOUNDS b = type_check(p, @\textcolor{type}{type[]}@);
\end{lstlisting}
\end{minipage} \\
\hline
{(e)} &
\begin{minipage}{0.93\columnwidth}
\begin{lstlisting}
type *p = &q->field; 
BOUNDS b = bounds_narrow(bq, q->field);
\end{lstlisting}
\end{minipage} \\
\hline
{(f)} &
\begin{minipage}{0.93\columnwidth}
\begin{lstlisting}
type *p = q + k; 
BOUNDS b = bq;
\end{lstlisting}
\end{minipage} \\
\hline
{(g)} &
\begin{minipage}{0.93\columnwidth}
\begin{lstlisting}[linebackgroundcolor=\highlightexx{\value{lstnumber}}]
bounds_check(p, b);
val = *p; $\mathit{or}$ *p = val; $\mathit{or}$ @\textrm{\texttt{p} escapes~\cite{duck16heap}}@
\end{lstlisting}
\end{minipage} \\
\end{tabular}
}
\caption{The dynamic type check instrumentation schema.
Here \texttt{b} represents the bounds for \texttt{p},
and \texttt{bq} for \texttt{q}.\label{fig:schema}}
\end{figure}

The next step is to ensure that all (derived) pointer
use is within the calculated bounds.
For this, rule Figure~\ref{fig:schema}(g) inserts a bounds check,
as represented by a call to a special (\verb+bounds_check+) function,
before each pointer use.
A call \verb+bounds_check(p,b)+ will report an error if:
{
\setlength\abovedisplayskip{4pt}
\setlength\belowdisplayskip{4pt}
\begin{align*}
\{\texttt{p}..\texttt{p}+\sizeof{\texttt{*p}}\} \cap b \neq
\{\texttt{p}..\texttt{p}+\sizeof{\texttt{*p}}\}
\end{align*}
}%
Rule Figure~\ref{fig:schema}(e) represents bounds \emph{narrowing}
to any sub-object selected by field access.
The (\verb+bounds_narrow+) operation between bounds (\verb+b+) and
field (\verb+p->field+)
is defined as interval intersection:
{
\setlength\abovedisplayskip{4pt}
\setlength\belowdisplayskip{4pt}
\begin{align*}
b \cap
\{(\texttt{\&p->field})..(\texttt{\&p->field}+\sizeof{\texttt{p->field}})\}
\end{align*}
}%
Narrowing is similar to that of MPX~\cite{mpx}.
Note that ordinary pointer arithmetic (e.g., array access) is not narrowed,
see rule~\ref{fig:schema}(f), since the resulting pointer may still
refer to the containing array.
Finally, we note that \tool will limit instrumentation to \emph{used}
pointers only (either directly or indirectly via a derived pointer).
For example, a function that merely casts and returns a pointer will not
attract instrumentation, unlike \caver, TypeSan, HexType and
\textsf{libcrunch}.
For \tool, it is the responsibility of the eventual user 
of the pointer to check the type.

\tool's instrumentation schema does not change the
\emph{Application Binary Interface} (ABI) nor does it rely on
disjoint mutable meta data to pass information between functions.
Instead, type/bounds information is always (re)calc\-ul\-at\-ed ``on demand''.
This helps maximize compatibility/thread-safety, which is essential when
instrumenting large code bases such as Firefox~\cite{firefox}.
However, this also assumes that input pointers are within the bounds of the
underlying object.
To help enforce this, rule~\ref{fig:schema}(g) also checks the
bounds of \emph{pointer escapes}
(e.g., passing a pointer as a parameter, writing a pointer to memory, etc.).
This is the same rationale used by \emph{low-fat pointers},
see~\cite{duck16heap} for more information.

\begin{example}[Dynamic Type Check Instrumentation]
Consider two functions:
(\texttt{length}) calculates the length of a linked-list and
(\texttt{sum}) calculates the sum of an array.
The instrumented versions of these
functions (using the Figure~\ref{fig:schema} schema) is shown in
Figure~\ref{fig:instrument}.
The instrumentation in lines $\{2,7,8,10,\-16,20\}$ is highlighted, and
the original functions can be obtained by deleting these lines and
eliminating temporary variables.
For the (\texttt{length}) function, the input pointer(s) (\texttt{xs})
on lines $\{2,10\}$ are type checked against the static type
(\type{node[]}) declared by the programmer.
This means that (\texttt{xs}) must point to an object (or sub-object of a
larger object) compatible with type (\type{node}).
The (\verb+type_check+) function does not guarantee (\texttt{xs}) points to
the base of a complete (\type{node}) object
(e.g., rule Figure~\ref{fig:layout}(b) allows (\texttt{xs}) to
point to the end of an object), so derived pointer (\texttt{tmp}) may be an
overflow.
To prevent this, the derived pointer (\texttt{tmp}) is bounds checked on
line 8.
Similarly, for the (\texttt{sum}) function, the input pointer (\texttt{a}) is
type checked against the static type (\type{int[]}), and the
derived pointer (\texttt{tmp}) is bounds checked before access.
Note how all pointer use (lines $\{9,21\}$) is preceded by a type/bounds check.
Figure~\ref{fig:instrument} also illustrates how the
number of type checks depends on the program itself.
For example,
(\texttt{length}) requires $O(N)$ type checks (one for each node in the list)
whereas (\texttt{sum}) only requires a single type check on function entry.
$\qed$
\end{example}\noindent
The (\texttt{sum}) function also highlights how the instrumentation
schema minimizes the number of type checks.
Here the input pointer (\texttt{a}) is type checked exactly once
outside of the loop,
whereas the subsequent derived pointers (\texttt{a+i}) are merely
bounds checked.

\begin{figure}
\vspace{2mm}
{\small
\newcommand{\highlightex}[1]{%
    \ifboolexpr{
        test {\ifnumcomp{#1}{=}{2}} or
        test {\ifnumcomp{#1}{=}{7}} or
        test {\ifnumcomp{#1}{=}{8}} or
        test {\ifnumcomp{#1}{=}{10}} or
        test {\ifnumcomp{#1}{=}{16}} or
        test {\ifnumcomp{#1}{=}{20}}
    }{\color{black!15}}{\color{row2}}
}
\lstset{language=C,
                frame=single,
                numbers=left,
                numberstyle=\scriptsize,
                xleftmargin=2em,
                xrightmargin=4pt,
                numbersep=6pt,
                basicstyle=\ttfamily\fontseries{m}\selectfont,
                deletekeywords={int},
                keywordstyle=\color{keyword}\ttfamily\fontseries{b}\selectfont,
                keywordstyle=[2]\color{type}\ttfamily\fontseries{b}\selectfont,
                keywordstyle=[3]\color{typecheck}\ttfamily\fontseries{b}\selectfont,
                stringstyle=\color{red}\ttfamily,
                commentstyle=\color{comment}\ttfamily\fontseries{b}\selectfont,
                morekeywords=[2]{void, TYPE, META, SUBOBJECT, BOUNDS, ssize_t,
                    size_t},
                morekeywords=[3]{type_check, type_match, type_error,
                    type_bounds, type_lookup, bounds_narrow, malloc,
                    lowfat_malloc, typed_malloc, bounds_check},
                linebackgroundcolor=\highlightex{\value{lstnumber}},
                mathescape
}
\begin{lstlisting}[escapechar=@]
int length(node *xs) {
   BOUNDS b = type_check(xs, @\textcolor{type}{node[]}@);
   int len = 0;
   while (xs != NULL) {
      len++;
      node **tmp = &xs->next;
      b = bounds_narrow(b, xs->next);
      bounds_check(tmp, b);
      xs = *tmp;
      b = type_check(xs, @\textcolor{type}{node[]}@);
   }
   return len;
}

int sum(int *a, int len) {
   BOUNDS b = type_check(a, @\textcolor{type}{int[]}@);
   int sum = 0;
   for (int i = 0; i < len; i++) {
      int *tmp = a + i;
      bounds_check(tmp, b);
      sum += *tmp;
   }
   return sum;
}
\end{lstlisting}
}
\caption{Instrumented \texttt{length} and \texttt{sum}
functions.\label{fig:instrument}}
\end{figure}

Finally, we remark that the Figure~\ref{fig:schema} schema is not designed
to be complete with respect to use-after-free errors.
For completeness, the combined type/bounds check and memory operation must be
atomic, else a call to (\texttt{free}), e.g., by another thread,
may mutate the type.
In practice, this means that some use-after-free errors may not be detected.
That said, complete use-after-free detection is not a design goal of \tool,
and even partial detection can be useful.
For example, \tool
detects known SPEC2006 use-after-free bugs (see Section~\ref{sec:experiments}).

\section{Dynamic Type Check Runtime}\label{sec:runtime}

\tool's runtime system is based on \emph{low-fat pointers}.

\mysubsubsection{Low-fat Pointers}
Low-fat pointers~\cite{duck16heap, duck17stack} are a method
for encoding bounds meta data (i.e., size and base of an allocation)
within the native machine pointer representation itself.
Low-fat pointers require sufficient pointer bit-width, and
are feasible for 64-bit systems (e.g., the \verb+x86_64+).
To use low-fat pointers, objects must be allocated using a special
\emph{low-fat memory allocator} that ensures the returned pointers are
suitably encoded.
The low-fat heap allocator~\cite{duck16heap} provides replacement functions
(\verb+lowfat_malloc+, \verb+lowfat_free+, etc.) to all the
\texttt{stdlib} memory allocation functions.
The replacement functions have the same interface (i.e., function prototype)
as the originals.
We also implement low-fat pointers for both stack~\cite{duck17stack} and
global~\cite{duck18extended} objects.

Several low-fat pointer encodings have been proposed.
For this paper, we use the low-fat pointer encoding
from~\cite{duck16heap, duck17stack}.
This encoding provides the following abstract operations:
given a low-fat pointer $p$ to (possibly the interior of) a low-fat allocated
object $O$, then:
{
\setlength\abovedisplayskip{4pt}
\setlength\belowdisplayskip{4pt}
\begin{align*}
\size{p} = \sizeof{O} & & \base{p} = \texttt{\&}O
\end{align*}
}%
That is, given a low-fat pointer $p$, we can use the $\mathit{size}$ and
$\mathit{base}$ operations to quickly determine the bounds meta data
of the allocated object.
E.g., if
{
\setlength\abovedisplayskip{4pt}
\setlength\belowdisplayskip{4pt}
\begin{align*}
\mathtt{str}=\mathtt{lowfat\_malloc}(\sizeof{\type{char[32]}})
\end{align*}
}%
then $\size{\mathtt{str}{+}10}{=}32$ and
$\base{\mathtt{str}{+}10}{=}\mathtt{str}$, etc.
Not all pointers will be low-fat pointers, and
such pointers are referred to as {\em legacy}.
For legacy pointer $q$,
$\size{q} = \texttt{SIZE\_MAX}$ and
$\base{q} = \texttt{NULL}$.
Support for legacy pointers is essential to handle
non-instrumented
code, e.g., libraries, and also some pointers 
from \emph{Custom Memory Allocators} (CMAs).

The low-fat pointer encoding of~\cite{duck16heap, duck17stack} works by
(1) arranging objects into different memory regions based on
allocation size, and (2) ensuring that all objects are allocation size-aligned.
Thus, given a pointer $p$, we can quickly derive the allocation size
(i.e., $\size{p}$) based on which memory region $p$ points into.
Next the $\base{p}$ operation is implemented by rounding $p$ down to the
nearest $\size{p}$-aligned address.
Both the $\size{p}$ and $\base{p}$ operations are fast and constant time
$O(1)$.
For more on low-fat pointers,
see ~\cite{duck16heap, duck17stack}.

\begin{figure}
\centering
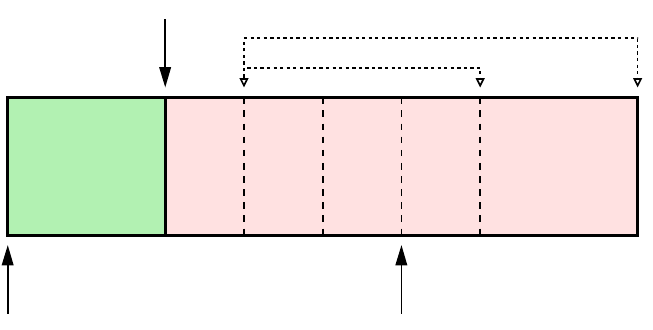
\caption{Object and object meta data layout.\label{fig:object}}
\end{figure}

\mysubsubsection{Using Low-fat Pointers for Meta Data}
Low-fat pointers were originally designed for \emph{allocation bounds
checking} using the meta data encoded in the pointer.
That is, given pointer $p$, any access outside the range
$\base{p} .. \base{p} {+} \size{p}$ is a bounds error that will
abort the program.
For \tool,
we repurpose low-fat pointers as a general method for
binding meta data (in our case, type information) to objects.
The basic idea is to store the meta data at the base of the object,
and this meta data can be retrieved from any interior pointer by
using the $\base{p}$ operation.
We refer to this as \emph{object meta data}, since it is associated with
every allocated object.
In the case of \tool, the object meta data contains a representation
of the dynamic type of the allocated object.
\begin{example}[Object Meta Data]\label{ex:object}
An example of the combined object and meta data layout is shown in
Figure~\ref{fig:object}.
Here we assume the object is of type (\type{T}) from
Example~\ref{ex:alloc_types}, and the layout of each sub-object
(e.g., \texttt{f}, \texttt{t}, \texttt{t.a[0]}, etc.) from
Example~\ref{ex:layout} is also illustrated.
The memory is divided into two main parts:
space for the object meta data (META) and
space for the allocated object itself.
Given a pointer $p$ to the object or sub-object
(e.g., $p{=}\texttt{\&t.a[2]}$ in Figure~\ref{fig:object}), the pointer
to the object meta data can be retrieved by $\base{p}$.
$\qed$\end{example}\noindent
It is important to note that the meta data (META) is
bound to the outermost object only, not each sub-object,
and occupies memory immediately
before the start of the object ($q$).
Thus, (META) is analogous to a ``malloc header'' that is
invisible to the program, and the layout of C/C++ objects is
otherwise unchanged.
Under our scheme, the (META) header is a type-integer pair
storing (1) the (top-level) allocation type of the object, and (2) the
object's allocation size (e.g., the parameter to (\texttt{malloc})).
For sub-objects, the type can be retrieved using the layout function
($\layoutF$) discussed below.

To implement the object layout of Figure~\ref{fig:object}, we
replace standard (\texttt{malloc}) with the version shown in
Figure~\ref{fig:runtime} lines 1-7.
Here, (\verb+type_malloc+) is a thin wrapper around the underlying
low-fat memory allocator (\verb+lowfat_malloc+).
The wrapper function takes a type (\texttt{t}) as an
argument (similar to C++'s \texttt{new} operator).
Here we treat types as first class objects of type (\type{TYPE}).
In lines 2-3, the wrapper allocates space for both the object
(of $\sizeof{\texttt{t}}$) and the object meta data
(of $\sizeof{\type{META}}$) using the underlying low-fat
allocator.
Lines 4-5 store the allocated object's meta data at the base address.
Line 6 returns the pointer to the start of the allocated object excluding
the meta data.
The (\verb+type_malloc+) function essentially binds the allocation type
(\texttt{t}) to the memory returned for the allocated object.
We similarly wrap low-fat
stack~\cite{duck17stack} and global~\cite{duck18extended}
objects with meta data.

Memory deallocation is handled by a (\verb+type_free+) 
replacement for stdlib (\texttt{free}).
The replacement function overwrites the object meta data with
the special type (\type{FREE}) defined in Section~\ref{sec:types}
before returning the memory to the underlying low-fat allocator.
The low-fat allocator has also been modified to ensure that the
meta data will be preserved until the memory is reallocated.
The (\verb+type_free+) function also detects double free errors.

\begin{figure}
\vspace{2mm}
{
\lstset{language=C,
                frame=single,
                numbers=left,
                numberstyle=\scriptsize,
                xleftmargin=2em,
                xrightmargin=4pt,
                numbersep=6pt,
                basicstyle=\small\ttfamily\fontseries{m}\selectfont,
                deletekeywords={int},
                keywordstyle=\color{keyword}\ttfamily\fontseries{b}\selectfont,
                keywordstyle=[2]\color{type}\ttfamily\fontseries{b}\selectfont,
                keywordstyle=[3]\color{typecheck}\ttfamily\fontseries{b}\selectfont,
                stringstyle=\color{red}\ttfamily,
                commentstyle=\color{comment}\ttfamily\fontseries{b}\selectfont,
                morekeywords=[2]{void, auto, TYPE, META, SUBOBJECT, BOUNDS, ssize_t,
                    size_t},
                morekeywords=[3]{type_check, type_match, type_error,
                    type_bounds, type_lookup, bounds_narrow, malloc,
                    lowfat_malloc, type_malloc},
                linebackgroundcolor={\ifodd\value{lstnumber}\color{row2}\fi},
                mathescape
}
\begin{lstlisting}[escapechar=@]
void *type_malloc(size_t size,TYPE t){
  META *meta = lowfat_malloc(
    sizeof(META)+size);
  meta$\rightarrow$type = t;
  meta$\rightarrow$size = size;
  return (void *)(meta + 1);
}

BOUNDS type_check(void *ptr, TYPE s) {
  META *meta = $\base{\texttt{ptr}}$;
  if (meta == NULL) /*legacy pointer*/
    return (0..UINTPTR_MAX);
  TYPE t = meta$\rightarrow$type;
  void *bptr = (void *)(meta + 1);
  BOUNDS b = (bptr..bptr+meta$\rightarrow$size);
  ssize_t k = ptr-bptr;
  for (auto o : $\layout{\texttt{t}}{\texttt{k}}$)
    if (o.type == s) {
      BOUNDS c = type_bounds(ptr, o);
      return bounds_narrow(b, c);
    }
  type_error();     /*report error*/
  return (0..UINTPTR_MAX);
}
\end{lstlisting}
}
\caption{Simplified definitions for the $\mathtt{type\_malloc}$ and
    $\mathtt{type\_check}$ functions.\label{fig:runtime}}
\end{figure}

\mysubsubsection{Type Checking with Meta Data}
By replacing the standard allocators, all objects
are bound to the allocation type which can be retrieved using the
$\mathit{base}$ operation.
Combined with the layout function ($\layoutF$),
the (\verb+type_check+) function can be implemented as shown in
Figure~\ref{fig:runtime} lines 9-24.
Here the (\verb+type_check+) function has three basic steps:
\begin{enumerate}
\item Get the allocation type (\texttt{t}), bounds (\texttt{b})
    and object base pointer (\texttt{bptr}) (lines 10-15);
\item Calculate the sub-object offset (\texttt{k}) (line 16);
\item 
    Scan all sub-objects at offset (\texttt{k}) as returned by the
    layout function ($\layoutF$) (lines 17-21).
    Return the bounds of the sub-object that matches the declared static
    type (\texttt{s}) narrowed to the allocation bounds,
    else raise a type error (line 22) if no match exists.
\end{enumerate}
For legacy pointers,
the (\verb+type_check+) function always succeeds and returns
``wide bounds'' 
(lines 11-12) for compatibility reasons.
Likewise, wide bounds are returned after a type error has been logged.

\begin{example}[Type Check]\label{ex:type_check}
Let $p$ point to an allocated object of type (\type{T}) from
Example~\ref{ex:layout}.
Assuming that $\sizeof{\type{META}}{=}16$,
the type will be stored as object meta data at address
$\base{p} = (p{-}\sizeof{\type{META}}) = (p{-}16)$.
Consider the interior pointer $(q{=}p{+}12)$.
Then $\mathtt{type\_check}(q, \type{int[]})$ computes:
\begin{enumerate}
\item $t = ((\type{META}~\texttt{*})\base{q})\texttt{->type} = \type{T}$
\item $k = q {-} \base{q} {+} \sizeof{\type{META}} = 12$
\item $\layout{\type{T}}{12} =
    \{\pair{\type{int[3]}}{8}, \pair{\type{int}}{0}, \pair{\type{int}}{4}\}$
\end{enumerate}
Type ($\type{int[]}$) matches the first sub-object
$\pair{\type{int[3]}}{8}$, and the bounds $p{+}4..p{+}16$ are returned.
On the other hand,
the $\mathtt{type\_check}(q, \type{double[]})$ will fail since there
is no matching sub-object for type (\type{double[]}).
$\qed$\end{example}\noindent
As illustrated in Example~\ref{ex:type_check}, it is sometimes possible to have
multiple matching sub-objects, i.e., $\pair{\type{int[3]}}{8}$,
$\pair{\type{int}}{0}$, and $\pair{\type{int}}{4}$ all match type (\type{int[]}).
In such cases, the following tie-breaking rules are used:
\begin{enumerate}
\item sub-objects with wider bounds are preferred; and
\item pointers-to-the-end-of-sub-objects (see Figure~\ref{fig:layout}(b))
    are matched last.
\end{enumerate}
Thus, the sub-object bounds for (\type{int[3]}) is returned.
Note that our approach for deriving (sub-)object bounds differs from that of
other systems such as SoftBound \cite{nagarakatte09softbound} and
MPX \cite{mpx}.
These systems track (sub-)object bounds by passing meta data whereas
\tool always (re)calculates bounds using the dynamic type.
Explicit tracking may allow for narrower bounds for some cases of type
ambiguity, e.g., when the bounds for (\type{int}) is intended.
In order to pass narrowed pointer arguments, SoftBound necessitates changing
the \emph{Application Binary Interface} (ABI).
\tool's approach achieves very good binary compatibility since the
underlying ABI is not changed.
Furthermore, SoftBound/MPX require meta data updates
when a pointer is written to memory, which creates a data race
for multi-threaded applications~\cite{oleksenko17mpx}.
Our approach requires no such updates,
allowing for better multi-threading support.

\mysubsubsection{Layout and Type Meta Data Implementation}
The object meta data is a representation of the dynamic
type of the corresponding object.
\tool represents incomplete types (i.e., \type{$T$[]})
as pointers to a
\emph{type meta data} structure containing useful information,
such as the type's size (i.e., $\sizeof{T}$), name (for
reflection) and layout information. 
The type meta data structure is defined once per type.

To reduce overheads, \tool uses a
\emph{layout hash table} representation.
The basic idea is as follows:
for all possible type (\type{T}), sub-object type (\type{S}) and
sub-object offset ($k$) combinations:
{
\setlength\abovedisplayskip{4pt}
\setlength\belowdisplayskip{4pt}
\begin{align*}
    \pair{\type{S}}{\delta} \in \layout{\type{T}}{k} & ~~~~ &
    0 \leq k \leq \sizeof{\type{T}}
\end{align*}
}%
the layout hash table will contain a corresponding entry:
{
\setlength\abovedisplayskip{4pt}
\setlength\belowdisplayskip{4pt}
\begin{align*}
    \type{T} \times \type{S} \times k \mapsto
    -\delta ~..~ \sizeof{\type{S}}{-}\delta
\end{align*}
}%
The entry maps a $(\type{T}{\times}\type{S}{\times}k)$ triple to the
corresponding sub-object bounds relative to offset $k$.
In order to keep the hash table finite, only entries corresponding to offsets
$0{\leq}k{\leq}\sizeof{\type{T}}$ are stored.
Otherwise, for entries outside this range, the offset is first
normalized $(k{:=}k \bmod \sizeof{\type{T}})$.
If multiple matching sub-objects exist for the same (\type{S})
the above tie-breaking rules apply.
Using this implementation, the sub-object matching of Figure~\ref{fig:runtime}
lines 17-21 can be efficiently implemented as an $O(1)$
hash table lookup.
\begin{example}[Layout Hash Table]
The layout hash table for (\type{T[]})
from Example~\ref{ex:layout} includes the following entries:
{
\setlength\abovedisplayskip{4pt}
\setlength\belowdisplayskip{4pt}
\begin{align*}
\begin{array}{c}
\hspace{-0.7em}
(\type{T}, \type{T}, 0) \mapsto -{\infty}..\infty \hspace{1em}
(\type{T}, \type{float} , 0) \mapsto 0..4 \hspace{1em}
(\type{T}, \type{S}, 4) \mapsto 0..20 \\
(\type{T}, \type{int}, 4) \mapsto 0..12 \hspace{1.5em}
(\type{T}, \type{int}, 8) \mapsto -4..8 \\
(\type{T}, \type{int}, 12) \mapsto -8..4 \hspace{1.5em}
(\type{T}, \type{char~*}, 16) \mapsto 0..8
\end{array}
\end{align*}
}%
Note that, since type (\type{T[]}) is incomplete,
the corresponding top-level entry is unbounded.\footnote{
The final bounds returned by (\texttt{type\_check}) is 
narrowed to the actual allocation size (Figure~\ref{fig:runtime} line 20)
stored in the object meta data.}
Consider the type check of pointer $(q{=}p{+}12)$ against 
(\type{int[]}) from Example~\ref{ex:type_check}.
The corresponding hash table entry
$(\type{T}, \type{int}, 12)$
maps to the bounds ($-8..4$).
Thus, the type check succeeds with the final bounds
$p{+}12{-}8..p{+}12{+}4 = p{+}4..p{+}16$.
Furthermore,
the type check of $q$ against (\type{double[]}) fails since there is no
corresponding layout hash table entry for $(\type{T}, \type{double}, 12)$.
$\qed$\end{example}\noindent
Our basic approach has also been extended to handle other standard C/C++
language features, including:
\begin{enumerate}
\item Structure types with \emph{flexible array members}; and
\item Automatic coercions between types
    allowable under the C, 
	``sloppy''~\cite{kell16type} or ``de facto''~\cite{memarian16depth}
    standards.
\end{enumerate}
Structures with \emph{Flexible Array Members} (FAMs) have
definitions of the form
$(\texttt{struct T \{$\cdots$; U member[];\}})$, where
(\verb+member+) of type (\type{U[]}) is the FAM.
Other forms are also possible.
The size of the FAM is determined by the object's allocation size.
Structures with a FAM are treated as
equivalent to
$(\texttt{struct T \{$\cdots$; U member[1];\}})$, and \tool uses an
alternative offset normalization for $k{>}\sizeof{\type{T}}$:
{
\setlength\abovedisplayskip{4pt}
\setlength\belowdisplayskip{4pt}
\begin{align*}
    k := ((k - \sizeof{\type{T}}) \bmod \sizeof{\type{U}}) + \sizeof{\type{T}}
\end{align*}
}%
The final feature is automatic coercion between different types,
such as
automatically coercing (\type{char[]}) to other types.
To implement this, \tool uses two layout hash table lookups instead of
one:
if the first lookup $(\type{T}, \type{S}, k)$ fails, next
$(\type{T}, \type{char}, k)$ is tried, representing
the coercion from (\type{char[]}) to (\type{S[]}).
This idea can be generalized to other kinds of useful coercions,
such as (\type{void *}) to (\type{S *}).

Type meta data, including the layout hash table, is automatically
generated using a compiler pass, once per compiled module.
Each type meta data object is declared as a \emph{weak symbol},
meaning that only one copy will be included in the final executable.
The type meta data is constant (read-only) and thus
cannot be modified at runtime.

\section{Experiments}\label{sec:experiments}

We have implemented a prototype version of \tool using the
LLVM compiler infrastructure~\cite{llvm} version 4.0.0 for the
\verb+x86_64+ architecture.
\tool's instrumentation is a two step process.
In the first step, a modified \texttt{clang} front-end generates a
\emph{type annotated} LLVM~\cite{llvm}
\emph{Intermediate Representation} (IR) of the C/C++ program.
Here,
type annotations are associated with each LLVM IR
instruction/global/function using the
standard DWARF~\cite{dwarf} debug format
(similar to that generated by the (\texttt{-g}) command-line option).
In the second step, the type annotated IR is instrumented 
using the schema from (Figure~\ref{fig:schema}).
This step also replaces all
heap/stack/global allocations with the typed variants
and generates the runtime type meta data
described in Section~\ref{sec:runtime}.
Our implementation supports all types described in
Section~\ref{sec:types}, including fundamental types, pointers, structures,
classes, unions, etc.,
as well as standard C/C++ features such as inheritance, virtual inheritance,
templates, multi-threading, basic coercions between
(\type{$T$}) to/from (\type{char[]}) and
(\type{$T$ *}) to/from (\type{void *}), and
objects with flexible array members.
In addition to the Figure~\ref{fig:schema} schema,
our \tool prototype supports basic optimizations such as:
removing dynamic type checks that can never fail
(e.g., C++ upcasts),
removing subsumed bounds checks,
and removing redundant bounds narrowing operations.
For speed, all instrumentation except (\verb+type_check+) is inlined.

By default, \tool logs all errors without stopping the program.
\tool may also be configured to merely count errors
(without detailed log messages),
and/or to abort after $N$ errors for some $N{\geq}1$.
For our experiments, logging mode is used to find errors,
and counting mode is used for measuring performance.

\mysubsubsection{Limitations}
The \tool prototype may not detect all possible type and memory errors.
For example, \tool can only partially protect legacy pointers in the form of
bounds narrowing.
For non-legacy pointers, \tool must correctly bind the
allocation type with each allocated object.
For global/stack objects as well as objects allocated using C++'s \verb+new+,
the allocation type is simply the declared type and is unambiguous.
However, for heap objects allocated by (\texttt{malloc})
we use a simple program analysis
(see Example~\ref{ex:alloc_types}).
Some \emph{Custom Memory Allocators} (CMAs)
use internal data structures to track memory, resulting in
type errors when cast.
For our experiments, we use a version of SPEC2006 with some CMAs removed,
see Appendix~\ref{sec:modifications}.
\tool will also not detect errors that are optimized away by LLVM
before the instrumentation pass.

For practical reasons,
the current prototype implements some simplifications, including:
treating \texttt{enum}s as (\type{int}),
C++ references as pointers, and
virtual function tables as arrays of generic functions.
Some simplifications are inherited from the
\texttt{clang} frontend.
The current prototype also does not aim to implement a strict
interpretation of the C/C++ standards.
For example, there is no tracking of pointer
\emph{provenance}~\cite{hathhorn15undef}.
Furthermore, the prototype implements some common
``sloppy'' \cite{kell16type} and ``de facto''~\cite{memarian16depth}
extensions,
such as (\type{T *}) to/from (\type{void *}) coercions.
The final limitation relates to sub-object matching.
By default, \tool heuristically chooses the sub-object with the widest
bounds (see the tie-breaking rules), which may 
differ from the intended bounds.
For example, given:
\begin{Verbatim}
    union { float a[10]; float b[20]; };
\end{Verbatim}
A type check against (\type{float[]}) will always return
\texttt{b}'s bounds.

\subsection{Effectiveness}\label{sec:effectiveness}

\begin{figure}[t]
\centering
{\small\setlength{\tabcolsep}{3pt}
\begin{tabular}{|l||r|rr|c|}
\hline
\multicolumn{5}{|c|}{SPEC2006} \\
\hline & kilo-~
       & \multicolumn{2}{l|}{~~checks (billions)}
       & \#Issues- \\
Bench. & sLOC & \#Type & \#Bound & found \\
\hline
\hline
\rowcolor{row}
\texttt{perlbench}     & 126.4 & 177.9  & 297.7  & 35  \\
\rowcolor{row}
\texttt{bzip2}         & 5.7   & 70.1   & 644.3  & 1   \\   
\rowcolor{row}
\texttt{gcc}           & 235.8 & 105.2  & 204.1  & 41  \\   
\texttt{mcf}           & 1.5   & 34.9   & 98.7   & 0   \\   
\texttt{gobmk}         & 157.6 & 90.9   & 421.3  & 0   \\   
\texttt{hmmer}         & 20.7  & 22.0   & 1393.4 & 0   \\   
\texttt{sjeng}         & 10.5  & 27.3   & 478.0  & 0   \\   
\texttt{libquantum}    & 2.6   & 276.4  & 561.1  & 0   \\       
\rowcolor{row}
\texttt{h264ref}       & 36.1  & 392.5  & 891.5  & 3   \\       
\texttt{omnetpp}$^\texttt{++}$
                       & 20.0  & 86.5   & 194.7  & 0   \\       
\texttt{astar}$^\texttt{++}$
                       & 4.3   & 72.5   & 216.8  & 0   \\   
\rowcolor{row}
\texttt{xalancbmk}$^\texttt{++}$
                       & 267.4 & 267.8  & 390.6  & 15  \\       
\hline
\rowcolor{row}
\texttt{milc}          & 9.6   & 29.4   & 347.1  & 1   \\   
\rowcolor{row}
\texttt{namd}$^\texttt{++}$
                       & 3.9   & 16.1   & 362.6  & 1   \\   
\rowcolor{row}
\texttt{dealII}$^\texttt{++}$    
                       & 94.4  & 266.1  & 701.3  & 13  \\   
\rowcolor{row}
\texttt{soplex}$^\texttt{++}$    
                       & 28.3  & 80.8   & 219.8  & 1   \\   
\rowcolor{row}
\texttt{povray}$^\texttt{++}$    
                       & 78.7  & 83.2   & 176.0  & 10  \\   
\rowcolor{row}
\texttt{lbm}           & 0.9   & 4.0    & 333.3  & 1   \\   
\rowcolor{row}
\texttt{sphinx3}       & 13.1  & 89.4   & 903.9  & 2   \\        
\hline
\hline
Totals (all)           & 1117.5 & 2193.0 & 8836.3 & 124 \\
\hline
Totals (C++)           & 497.0  & 873.1  & 2261.7 & 40  \\
\hline
\end{tabular}
}
\caption{Summary of the SPEC2006 benchmarks.
    C++ benchmarks are marked with a (\texttt{++}), and the rest are C.
    We bucket issues by type and offset.
    Benchmarks with issues are highlighted.\label{fig:summary}}
\vspace{-3mm}
\end{figure}

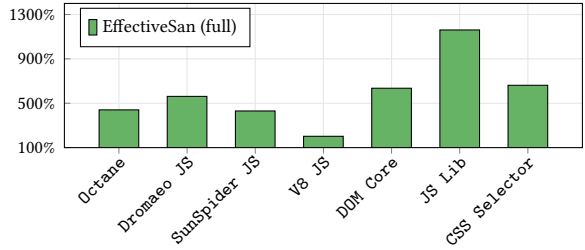
\begin{figure}
\pgfplotsset{
    axisA/.style={
        ybar=0pt,
        ymin=100,
        ymax=1400,
        xticklabels={
           \texttt{\scriptsize Octane},
           \texttt{\scriptsize Dromaeo JS},
           \texttt{\scriptsize SunSpider JS},
           \texttt{\scriptsize V8 JS},
           \texttt{\scriptsize DOM Core},
           \texttt{\scriptsize JS Lib},
           \texttt{\scriptsize CSS Selector},
        },
        ytick={100,500,900,1300},
        yticklabels={
            {\scriptsize 100\%},
            {\scriptsize 500\%},
            {\scriptsize 900\%},
            {\scriptsize 1300\%},
        },
        bar width=15pt,
        x tick label style={rotate=45,anchor=east,yshift=-2.5pt,xshift=3pt},
        width=\columnwidth,
        height=3.5cm,
        xtick=data,
        xtick pos=left,
        ytick pos=left,
        legend cell align={left},
        legend pos=north west,
        major tick length=0.08cm,
        enlarge x limits={true, abs value=0.8},
        grid style={gray!20},
        grid=both,
        legend image code/.code={%
            \draw[#1] (0cm,-0.1cm) rectangle (3pt,3pt);
        },
        title style={yshift=-6pt}
        }
}
\pgfplotstableread{
0    440.251775338928
1    561.893280218052
2    430.441003350539
3    202.108059082664
4    635.427706283119
5    1160.77057793345
6    662.033532906068
}\dataset
\begin{tikzpicture}
    \begin{axis}[
        axisA
      ]
      \addplot[fill=mygreen!60] table[x index=0,y index=1] \dataset;
      \legend{{\scriptsize \tool (full)}};
    \end{axis}
\end{tikzpicture}
\vspace{-3mm}
\caption{\tool relative performance for Firefox and
    various standard browser benchmarks.\label{fig:firefox}}
\end{figure}

To test the effectiveness of \tool, we use the 
SPEC2006 \cite{henning06spec} benchmarks
and the Firefox web browser version 52 (ESR).
The SPEC2006 benchmarks ($\sim$1.1million sLOC) comprise several
integer and floating point C/C++ programs.
For SPEC2006 we use the standard workloads.
For Firefox ($\sim$7.9million sLOC) we use standard web benchmarks
(see Figure~\ref{fig:firefox}).

The results for SPEC2006 are summarized in Figure~\ref{fig:summary}.
Here (kilo-sLOC) represents the source lines of code (in thousands),
(\#Type) the number of type checks,
(\#Bounds) the number of bounds checks,
(\#Issues-found) the number of issues logged by \tool.
We bucket issues by type and offset to prevent the same issue
from being reported at multiple different program points.
Of the $\sim$2.2 trillion type checks in Figure~\ref{fig:summary}, only
$\sim$1.1\% were performed on legacy pointers,
meaning that \tool achieves high coverage.

For SPEC2006 our \tool prototype detects several issues
(see Figure~\ref{fig:summary}), including:
\begin{itemize}[leftmargin=*]
\item A use-after-free bug in \texttt{perlbench}
(reported in~\cite{serebryany12asan}).\footnote{
Only applicable to the SPEC2006 \texttt{test} workload.}
\item A bounds overflow error in \texttt{h264ref}
(reported in~\cite{serebryany12asan}).
\item Three sub-object bounds overflow errors in
    \texttt{gcc}, \texttt{h264ref} and \texttt{soplex}.\footnote{
Some are also found by MPX, see~\cite{oleksenko17mpx}.
}
\item Multiple type errors (discussed below).
\end{itemize}
As far as we are aware, 
all previously known bounds,
type confusion, and use-after-free errors are detected.
\tool also detects new type errors that have not been
previously reported (see below).

Interestingly, \tool reports zero issues (on executed paths)
for the \texttt{mcf},
\texttt{gobmk}, \texttt{hmmer}, \texttt{sjeng}, \texttt{libquantum},
\texttt{om\-net\-pp}, and \texttt{astar} benchmarks.
Similarly, the benchmarks \texttt{bzip2}, \texttt{h264ref},
\texttt{milc}, \texttt{namd}, \texttt{soplex}, \texttt{lbm} and
\texttt{sphinx3} report one or two minor issues.
This shows that it is feasible for well disciplined C/C++ code to have zero
type and memory errors.
Of the remaining benchmarks, \texttt{perlbench}, \texttt{gcc}, and
\texttt{povray} had the most issues, as is discussed below.
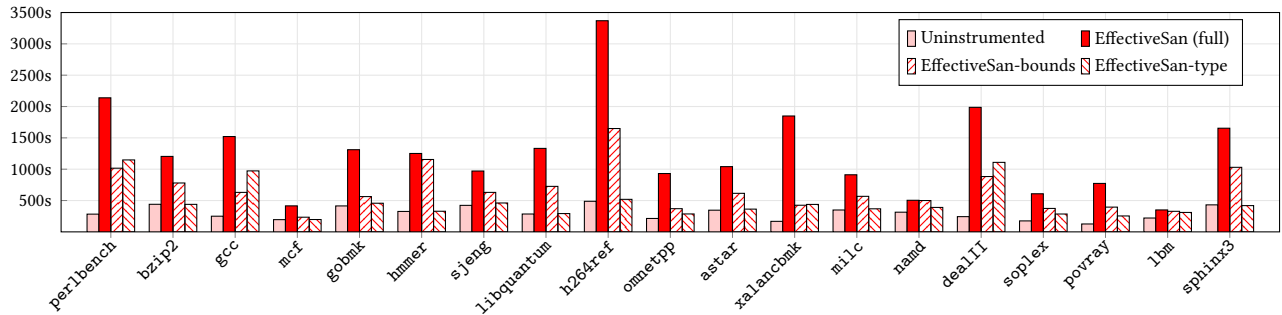
\begin{figure*}
\pgfplotsset{
    axisA/.style={
        ybar=0pt,
        ymin=0,
        ymax=3500,
        xticklabels={
           \texttt{\scriptsize perlbench},
           \texttt{\scriptsize bzip2},
           \texttt{\scriptsize gcc},
           \texttt{\scriptsize mcf},
           \texttt{\scriptsize gobmk},
           \texttt{\scriptsize hmmer},
           \texttt{\scriptsize sjeng},
           \texttt{\scriptsize libquantum},
           \texttt{\scriptsize h264ref},
           \texttt{\scriptsize omnetpp},
           \texttt{\scriptsize astar},
           \texttt{\scriptsize xalancbmk},
           \texttt{\scriptsize milc},
           \texttt{\scriptsize namd},
           \texttt{\scriptsize dealII},
           \texttt{\scriptsize soplex},
           \texttt{\scriptsize povray},
           \texttt{\scriptsize lbm},
           \texttt{\scriptsize sphinx3}
        },
        ytick={500,1000,1500,2000,2500,3000,3500,4000},
        yticklabels={
            {\scriptsize 500s},
            {\scriptsize 1000s},
            {\scriptsize 1500s},
            {\scriptsize 2000s},
            {\scriptsize 2500s},
            {\scriptsize 3000s},
            {\scriptsize 3500s},
            {\scriptsize 4000s}
        },
        bar width=4.5pt,
        x tick label style={rotate=45,anchor=east,yshift=-2.5pt,xshift=3pt},
        width=\textwidth,
        height=4.5cm,
        xtick=data,
        xtick pos=left,
        ytick pos=left,
        legend cell align={left},
        legend pos=north east,
        major tick length=0.08cm,
        enlarge x limits={true, abs value=0.8},
        grid style={gray!20},
        grid=both,
        legend image code/.code={%
            \draw[#1] (0cm,-0.1cm) rectangle (3pt,3pt);
        },
        legend columns=2,
        title style={yshift=-6pt}
        }
}
\pgfplotstableread{
0   283    2139 1016 1148 
1   440    1204 780  439 
2   250    1521 630  973
3   195    415  234  197 
4   414    1311 563  457
5   326    1251 1155 329
6   423    971  630  461 
7   285    1332 726  292
8   488    3369 1649 519 
9   214    931  370  286 
10  346    1041 615  363 
11  168    1849 425  438 
12  349    911  567  368 
13  314    504  500  390 
14  243    1987 883  1109 
15  175    608  374  285 
16  126    773  395  254 
17  221    350  328  310 
18  431    1654 1030 419 
}\dataset
\begin{tikzpicture}
    \begin{axis}[
        axisA
      ]
      \addplot[fill=red!20] table[x index=0,y index=1] \dataset;
      \addplot[fill=red]  table[x index=0,y index=2] \dataset;
      \addplot[pattern color=red, pattern=north east lines]  table[x index=0,y index=3] \dataset;
      \addplot[pattern color=red, pattern=north west lines]  table[x index=0,y index=4] \dataset;
      \legend{{\scriptsize Uninstrumented},{\scriptsize \tool (full)}, 
        {\scriptsize \tool-bounds}, {\scriptsize \tool-type}};
    \end{axis}
\end{tikzpicture}
\vspace{-3mm}
\caption{\tool SPEC2006 timings in seconds.
We test three \tool variants:
    \tool (full instrumentation),
    \tool-bounds (object bounds checking only), and
    \tool-type (type cast checking only).
\label{fig:timings}}
\end{figure*}

\mysubsubsection{Type Errors}
\tool detects multiple type errors in the SPEC2006 benchmarks, including:
\begin{itemize}[leftmargin=*]
\item[-] Bad C++ downcasts (type confusion) in \texttt{xal\-anc\-bmk};
\item[-] Multiple instances of casting to container types, i.e.:
    (\type{T}) cast to (\type{struct S \{T t; $...$ \}}) for some
    \type{T}, \type{S}.
    Several instances relating to \verb_stdlib++_ are similar to those
    previously reported by \caver~\cite{lee15caver}.
\item[-] Multiple instances of casting between classes/structures
    with the same layout (e.g., phantom classes~\cite{lee15caver}).
\item[-] \texttt{gcc}/\texttt{sphinx3} casts objects to (\type{int[]}) to
calculate hash values or checksums.
\item[-] \texttt{gcc} with incompatible definitions for the
same type
(e.g., different \texttt{struct} definitions using the same tag).
\item[-] \texttt{bzip2}/\texttt{lbm} confuses fundamental types
(\texttt{lbm} case also reported in~\cite{ireland13safetype}).
\item[-] \texttt{perlbench} reusing memory (as a different type) rather than
explicitly freeing it.
\item[-] \texttt{perlbench} frequently confuses (\type{$T$ *}) with
    (\type{$T$ **}).
\item[-] \texttt{perlbench}/\texttt{povray}'s ad hoc implementation of
C++-style inheritance by defining structures with a common shared prefix,
and casting to and fro.
\end{itemize}
Two type errors from the \texttt{xal\-anc\-bmk} benchmark relate to
bad C++ downcasting, similar to the class of errors
detectable by \caver, TypeSan and HexType.
The first arises from:
\begin{Verbatim}
SchemaGrammar& sGrammar =
  (SchemaGrammar&) grammarEnum.nextElement();
\end{Verbatim}
This operation represents a downcast from the base class (\type{Grammar}) 
(returned by \texttt{nextElement})
to the derived class (\type{SchemaGrammar}).
However, at runtime, \texttt{nextElement} may also return
a (\type{DTDGrammar}), which is neither a base nor derived class of
(\type{SchemaGrammar}), and thus the downcast is invalid.
A second case arises from an invalid downcast
from a (\type{DOMDocumentImpl}) to a (\type{DOMElementImpl}).
In both cases, the result of the bad cast is only used to access virtual
methods from a shared base class.
The code relies on undefined behavior, namely, that
the virtual function tables of the derived classes are
compatible.

Some type errors relate to \emph{type abuse}, i.e., likely deliberate
type errors introduced by the programmer.
For example, \texttt{perlbench} and \texttt{povray} use an idiom that confuses
\texttt{struct}s with shared common prefixes as an ad hoc implementation of
C++-style inheritance, e.g.
\begin{Verbatim}
 struct Base { int x; float y; };
 struct Derived { int x; float y; char z; };
\end{Verbatim}
The (\type{Base}) and (\type{Derived}) structure types are incompatible
(\cite{c} \S 6.2.7),
thus accessing an object of one type through the other is
undefined behavior (\cite{c} \S 6.5.0 \P 7).
Such idioms may break the compiler's \emph{Type-Based Alias Analysis}
(TBAA)~\cite{diwan98tbaa} assumptions
and cause programs to be mis-compiled---a known problem for
\texttt{perlbench}~\cite{henning06spec}.
The code can be re-factored as follows to avoid type errors:
\begin{Verbatim}
 struct Derived { struct Base base; char z; };
\end{Verbatim}
Alternatively \texttt{union}s or standard C++ \texttt{class}es with
inheritance can be used.

\mysubsubsection{Memory Errors}
In addition to previously reported~\cite{serebryany12asan} memory errors
in \texttt{perlbench} and \texttt{h264ref},
\tool detects the following sub-object bounds overflows:
\begin{itemize}[leftmargin=*]
\item[-] \texttt{gcc} overflows the (\texttt{mode}) field of
    type (\verb+rtx_const+) to access structure padding inserted
    by the compiler.
\item[-] \texttt{h264ref} overflows the (\verb+blc_size+) field of an
    object of type (\type{InputParameters}).
\item[-] \texttt{soplex} \underline{underflows} the (\verb+themem1+) field of
    an object of type (\type{UnitVector}).
\end{itemize}
The \texttt{soplex} underflow appears to be intentional
(it is documented in the source code comments),
and relies on the compiler not inserting padding between fields.

Interestingly, the \texttt{gcc} error is not
reported by MPX~\cite{oleksenko17mpx}.
This is possibly because MPX assumes the static type (\type{int[]}) is correct
and does not narrow.
In contrast, \tool matches the static type against the first field of
dynamic type (\verb+rtx_const+), implying much narrower bounds.
Furthermore, \tool does not report false positives that affect other tools.
For example,
\texttt{xalancbmk} performs
container-style subtraction from the
base of a structure, which is reported as a sub-object bounds
overflow by MPX~\cite{oleksenko17mpx}.
However, this is not considered a sub-object overflow by \tool, since
the operation involves a cast to (\type{char *}), resetting the bounds to
the containing object.

\mysubsubsection{Discussion}
As noted above, some issues found by \tool correspond to intentional
type/memory abuse introduced by the programmer, and not unintentional bugs.
Even memory errors,
such as the \texttt{soplex} sub-object bounds underflow (detailed above),
may be intentional.
\tool does not currently distinguish between
intentional abuse and unintentional bugs, as
such a distinction relies on
application rather than language semantics,
and is therefore best left to the programmer.

That said, even exposing intentional type/memory abuse can be useful,
such as for code quality or standards compliance reasons.
Removing abuse may also help isolate more serious issues.
For example, \texttt{perl\-bench} is rife with type
error abuse---resulting in a large error log---which makes finding
``real'' bugs more difficult.
Type errors may also identify opportunities for code refactoring.
For example, \tool detects multiple type errors in the \texttt{povray}
benchmark
relating to an idiosyncratic implementation of C++-style inheritance using
C-style \texttt{struct}s with overlapping layouts.
This was surprising, given that \texttt{povray} itself is
implemented in C++,
and is possibly an artifact of a previous C to C++ transition.
Such idioms also affect code quality, since the 
object hierarchy is represented in a non-standard way,
affecting code readability.
The \texttt{povray} type errors can be resolved 
by switching to standard C++ \texttt{class}es and inheritance.
Indeed, modern versions of \texttt{povray}\footnote{
\url{http://www.povray.org}} have done so.
Finally, some type errors may help identify 
\emph{Custom Memory Allocators} (CMAs) used by the program.
Such CMAs can be replaced with standard allocators
to help improve the accuracy of \tool and related tools, as was done with
the SPEC2006 benchmarks.

\subsection{Performance}

\mysubsubsection{Timings}
To evaluate performance, we test the \tool prototype against
the SPEC2006 benchmark suite~\cite{henning06spec}.
All experiments are run on a Xeon E5-2630v4 processor (clocked at 2.20GHz)
with 32GB of RAM.
The results are shown in Figure~\ref{fig:timings}.
Here we evaluate three different variants:
\begin{itemize}[leftmargin=*]
\item \tool (full): full \tool instrumentation;
\item \tool-bounds: protects object bounds only;
\item \tool-type: protects bad C/C++ casts only.
\end{itemize}
\tool-bounds protects object bounds only
by replacing type check
instrumentation (rules Figure~\ref{fig:schema}(a)-(d)) with a simpler
(\verb+bounds_get+) function that returns the allocated object bounds
without checking whether the type is correct.
The object bounds are calculated from the object's dynamic type $T$,
i.e., by $\sizeof{T}$.
\tool-type restricts type checking to C/C++-style cast operations only
(rule Figure~\ref{fig:schema}(d)) and
all other instrumentation is removed.
Unlike \tool (full), rule \ref{fig:schema}(d) is applied regardless of
whether the resulting pointer is used.
The main motivation for the variants is to enable a meaningful
comparison with related tools such as
AddressSanitizer and HexType.
We also compare against the uninstrumented baseline at
(\texttt{-O2}).

The additional performance overheads of \tool, \tool-bounds and \tool-type are
288\%, 115\% and 49\% respectively (Figure~\ref{fig:timings}).\footnote{
The protection and overhead of each \tool variant
shown in Figure~\ref{fig:timings} is not meant to be additive.
For example, only full \tool can detect non-cast type errors 
and sub-object bounds overflow errors.}
Unsurprisingly, \tool with full comprehensive instrumentation enabled has
the highest overhead, at 288\% overall.
However, this mode makes no assumptions about the type and bounds of objects,
so is the most likely to find errors in the program.
Reducing the instrumentation trades error coverage for performance, as
demonstrated by the 
(\tool-bounds) and (\tool-type) variants,
with runtime overheads of 115\% and 49\% respectively.

\mysubsubsection{Tool Comparison}
To understand the cost of bounds checks, \tool-bounds
can be compared to more specialized sanitizers that protect object or
allocation bounds only, 
such as AddressSanitizer (73\%~\cite{serebryany12asan}, or 92\% reported
in~\cite{duck17stack}), 
SoftBound (67\%~\cite{nagarakatte09softbound} for partial
SPEC2000/SPEC2006;
between 60-249\% for 4/19 SPEC2006~\cite{kuznetsov14code};
$\sim$100\% for 6/19 SPEC2006~\cite{oleksenko17mpx}),
MPX ($\sim$200\% for SPEC2006~\cite{oleksenko17mpx}),
BaggyBounds (60\%~\cite{akritidis09baggy} for SPEC2000),
and LowFat (54\%~\cite{duck17stack} for SPEC2006).
The overhead of \tool-bounds is higher than most tools,
but is not meant to replace specialized solutions.
\tool's meta data representation
is primarily designed for type checking, meaning that object bounds must be
calculated indirectly from dynamic type information.

\tool-type's instrumentation is comparable to existing
type confusion sanitizers such as \caver~\cite{lee15caver},
TypeSan~\cite{haller16typesan} and HexType~\cite{jeon17hextype}.
\caver reports a 20.0-29.6\% overhead
for 2/19 SPEC2006 benchmarks, 
TypeSan a 12.1\% overhead for 7/19 benchmarks, and
HexType a 3.3\% overhead for 7/19 benchmarks.
\tool-type has higher overhead, at 49\% for all of SPEC2006.
However, these existing sanitizers are specialized for casts
between C++ \texttt{class}es, which results in far less checking.
For example, TypeSan does a total of 5.9 billion
type checks for all SPEC2006 C++ benchmarks~\cite{haller16typesan},
whereas \tool (full) does 873.1 billion
(Figure~\ref{fig:summary}, excluding bounds checks)
and \tool-type does 361.1 billion, with
264.2 billion from \texttt{perlbench}, \texttt{gcc}, and \texttt{dealII} alone.
These existing sanitizers do not handle C programs
(\texttt{perlbench} and \texttt{gcc})
nor C-style casts from \texttt{dealII}, which account for most of the
additional checks.
Our results also show that the overhead-per-check ratio strongly favors
\tool.
HexType~\cite{jeon17hextype} also uses optimizations
such as avoiding tracking for objects that are never cast.
In principle, such optimizations could also be adapted to \tool-type,
however \tool-type is not meant to replace specialized tools.
Finally, we note that
\tool is also significantly faster than previous runtime type checking systems
for C, such as~\cite{loginov01typecheck} with a
35$\times$-133$\times$ slowdown for SPEC95.

\begin{figure}
\pgfplotsset{
    axisA/.style={
        ybar=0pt,
        ymin=0,
        ymax=2200,
        xticklabels={
           \texttt{\scriptsize perlbench},
           \texttt{\scriptsize bzip2},
           \texttt{\scriptsize gcc},
           \texttt{\scriptsize mcf},
           \texttt{\scriptsize gobmk},
           \texttt{\scriptsize hmmer},
           \texttt{\scriptsize sjeng},
           \texttt{\scriptsize libquantum},
           \texttt{\scriptsize h264ref},
           \texttt{\scriptsize omnetpp},
           \texttt{\scriptsize astar},
           \texttt{\scriptsize xalancbmk},
           \texttt{\scriptsize milc},
           \texttt{\scriptsize namd},
           \texttt{\scriptsize dealII},
           \texttt{\scriptsize soplex},
           \texttt{\scriptsize povray},
           \texttt{\scriptsize lbm},
           \texttt{\scriptsize sphinx3}
        },
        ytick={500,1000,1500,2000},
        yticklabels={
            {\scriptsize 500MB},
            {\scriptsize 1000MB},
            {\scriptsize 1500MB},
            {\scriptsize 2000MB}
        },
        bar width=3.5pt,
        x tick label style={rotate=45,anchor=east,yshift=-2.5pt,xshift=3pt},
        width=\columnwidth,
        height=3.5cm,
        xtick=data,
        xtick pos=left,
        ytick pos=left,
        legend cell align={left},
        major tick length=0.08cm,
        enlarge x limits={true, abs value=0.8},
        grid style={gray!20},
        grid=both,
        legend image code/.code={%
            \draw[#1] (0cm,-0.1cm) rectangle (3pt,3pt);
        },
        title style={yshift=-6pt}
        }
}
\pgfplotstableread{
0   680    674 
1   872    848  
2   908    868
3   1718   1677  
4   31     36     
5   28     32     
6   180    177    
7   100    97   
8   67     69    
9   175    199     
10  335    396  
11  432    554   
12  697    688     
13  49     48     
14  815    1028    
15  443    588     
16  7      7     
17  420    410     
18  46     50     
}\dataset
\begin{tikzpicture}
    \begin{axis}[
        axisA
      ]
      \addplot[fill=blue!20] table[x index=0,y index=1] \dataset;
      \addplot[fill=blue]  table[x index=0,y index=2] \dataset;
      \legend{{\scriptsize Uninstrumented},{\scriptsize \tool (full)}};
    \end{axis}
\end{tikzpicture}
\vspace{-3mm}
\caption{\tool (full) memory usage in MB.\label{fig:memory}}
\end{figure}

\mysubsubsection{Memory}
The memory overheads (peak resident set size) for \tool are shown in
Figure~\ref{fig:memory}.
Overall we see that \tool introduces a ${\sim}12\%$ memory overhead,
which is comparable to the ${\sim}3\%$ overhead 
introduced by the underlying low-fat pointer implementation~\cite{duck17stack}.
This also suggests that the memory overheads introduced by object and
type meta data are modest.
Many existing sanitizers that use shadow memory report higher memory
overheads, e.g., AddressSanitizer~\cite{serebryany12asan} at 237\% overhead.

\subsection{Web Browser Evaluation}
We evaluate \tool against Firefox~\cite{firefox} in order to test complex and 
multi-threaded software.
Firefox is built using \tool after:
(1) disabling \texttt{jemalloc}\footnote{
Disabling \texttt{jemalloc} is also standard practice for compiling Firefox
with AddressSanitizer.
}
(2) replacing components written directly in assembly
(\tool assumes C/C++ source code), and
(3) applying
a one-line patch that removed stack object ordering assumptions
that are incompatible with the low-fat stack allocator~\cite{duck17stack}.
Aside from \texttt{jemalloc}, we instrument Firefox ``as is'' without
replacing other \emph{Custom Memory Allocators} (CMAs),
the same approach used by~\cite{haller16typesan}.
Finally, we note that, as far as we are aware, \tool is the first full
type and sub-object bounds checker used to build a web browser,
demonstrating the compatibility of our approach.
Other sub-object bounds checkers, such as MPX and SoftBound,
do not support multi-threaded code~\cite{oleksenko17mpx}
required for browsers.

The results for standard browser benchmarks are shown in
Figure~\ref{fig:firefox}.
Overall we see that \tool (full) introduces a 422\% overhead compared to the
uninstrumented baseline, which is (1.5$\times$) the additional overhead
compared to the SPEC2006 results.
Although the overhead for Firefox is higher,
our result is consistent with \caver
(2.6$\times$ for 2/19 SPEC2006)~\cite{lee15caver},
TypeSan (2.8$\times$ for 7/19 SPEC2006)~\cite{haller16typesan},
and HexType (55$\times$ for 7/19 SPEC2006)~\cite{jeon17hextype} which
similarly report higher overheads for Firefox relative to the SPEC2006
benchmarks.
In~\cite{haller16typesan} it is noted that Firefox creates large
numbers of temporary objects which leads to increased overheads
for tools implementing type checking.

\tool detects multiple issues for Firefox summarized below.
Most issues relate to \emph{type abuse} (similar to our SPEC2006 results) or
CMAs, including:
\begin{itemize}[leftmargin=*]
\item[-] Multiple instances of casts between types that are
    equivalent modulo template parameters.
    For example, an
    object of type (\type{$T$<$U$*>}) being cast to (\type{$T$<void*>}) and
    vice versa,
    such as
    (\type{nsTArray\_Impl<void*>}) being confused with
    (\type{nsTArray\_Impl<PVRLayerParent*>}),
    etc.
\item[-] Multiple instances of type abuse similar to our
    SPEC2006 results, including:
    casting to container types and casting structures to
    fundamental types (e.g., \type{int[]}).
\item[-] Multiple errors relating to the use of CMAs.
    For example, function (\texttt{XPT\_ArenaCalloc}) is one such CMA that
    returns objects typed with an internal allocator
    structure (\type{BLK\_HDR}).
    This results in type errors, e.g.,
    (\type{BLK\_HDR}) versus
    (\type{XPTMethodDescriptor}), etc.
\end{itemize}
The latter demonstrates how type errors can sometimes 
identify CMAs.
Such CMAs can be replaced with standard
allocators to better assist dynamic analysis tools such as \tool.
However, due to the size and complexity of the Firefox code-base,
such an exercise is left as future work.

\section{Conclusion}\label{sec:conclusion}

In this paper, we have proposed dynamic typing as a general method for
comprehensive type and memory error detection in C/C++ programs.
We also presented \tool, a practical implementation of dynamic typing
using a combination of low-fat pointers, meta data, and
type/bounds check in\-tru\-ment\-at\-ion.
We have evaluated \tool against the SPEC2006 benchmark suite and
Firefox, finding several new errors.
We also show that \tool is effective at detecting sub-object bounds
errors, one of only a few tools that can do so, while being
compatible with multi-threaded environments and preserving the Application
Binary Interface.

The scope for future work is broad.
\tool's method for tracking dynamic type information can likely be
generalized to other useful properties, enabling
new classes of C/C++ sanitizers.
The performance of our prototype can also likely be improved as
new optimizations are implemented.

\appendix
\section{SPEC2006 Modifications}\label{sec:modifications}

For our SPEC2006 experiments, the following CMAs/wrappers were replaced with
the standard (\verb+malloc+) equivalent:

{\small
\begin{Verbatim}[commandchars=\\\{\}]
     Perl_malloc\textrm{,} safemalloc\textrm{,} Perl_safesysmalloc\textrm{,}
      BZALLOC\textrm{,} xmalloc\textrm{,} pov_malloc\textrm{,} MallocOrDie\textrm{,}
   MemoryManager::allocate\textrm{,} XMemory::operator new\textrm{,}
           __ckd_malloc__\textrm{,} __mymalloc__
\end{Verbatim}
}\noindent
The analogous CMAs/wrappers for (\verb+realloc+), (\verb+calloc+) and
(\verb+free+) were also replaced.

\newpage

\bibliography{types}

\end{document}